\documentclass[prb,twocolumn,superscriptaddress,amsmath,amssymb,floatfix,preprintnumbers]{revtex4}
\usepackage{epsfig}
\usepackage{amsmath}
\usepackage{graphicx}
\usepackage{placeins}
\usepackage{float}
\usepackage{dcolumn}
\usepackage{color}
\usepackage{lipsum}
\usepackage{epsfig}  
\usepackage{epstopdf}
\usepackage{physics} 
\usepackage{bm}
\usepackage[all]{xy}
\usepackage{braket}

\DeclareGraphicsExtensions{.png .jpg .pdf}

\usepackage{dsfont}

\usepackage{mathtools} 
\graphicspath{{figure/}}
\usepackage{hyperref}
\hypersetup{
     colorlinks = true,
     linkcolor = blue,
     anchorcolor = blue,
     citecolor = blue,
     filecolor = blue,
     urlcolor = blue
     }

\begin{document}

\title{Electronic band evolution between Lieb and kagome nanoribbons}

\author{E. S. Uchôa}
\thanks{These two authors contributed equally}
\email{emiliauchoa@fisica.ufc.br}
\affiliation{Departamento de F\'isica, Universidade Federal do Cear\'a, Campus do Pici, 60455-900, Fortaleza, Cear\'a, Brazil}

\author{W. P. Lima}\thanks{These two authors contributed equally}\email{wellisson@fisica.ufc.br}
\affiliation{Departamento de F\'isica, Universidade Federal do Cear\'a, Campus do Pici, 60455-900, Fortaleza, Cear\'a, Brazil}

\author{S. H. R. Sena}\email{silviahelena@unilab.edu.br}
\affiliation{Instituto de Ci\^encias Exatas e da Natureza, Universidade da Integra\c{c}\~ao Internacional da Lusofonia Afro-Brasileira, Campus das Auroras, 62790-970, Reden\c{c}\~ao, Cear\'a, Brazil}

\author{A. J. C. Chaves}\email{andreJ6@gmail.com}
\affiliation{Departamento de F\'isica, Instituto Tecnol\'ogico de Aeron\'autica,  12228-90, S\~ao Jos\'e dos Campos, S\~ao Paulo, Brazil}

\author{J. M. Pereira Jr.}\email{pereira@fisica.ufc.br}
\affiliation{Departamento de F\'isica, Universidade Federal do Cear\'a, Campus do Pici, 60455-900, Fortaleza, Cear\'a, Brazil}

\author{D. R. da Costa}\email{diego\_rabelo@fisica.ufc.br}
\affiliation{Departamento de F\'isica, Universidade Federal do Cear\'a, Campus do Pici, 60455-900, Fortaleza, Cear\'a, Brazil}
\affiliation{Department of Physics, University of Antwerp, Groenenborgerlaan 171, B-2020 Antwerp, Belgium}

\begin{abstract}
We investigate the electronic properties of nanoribbons made out of monolayer Lieb, transition, and kagome lattices using the tight-binding model with a generic Hamiltonian. It allows us to map the evolutionary stages of the interconvertibility process between Lieb and kagome nanoribbons by means of only one control parameter. Results for the energy spectra, the density of states, and spatial probability density distributions are discussed for nanoribbons with three types of edges: straight, bearded, and asymmetric. We explore for different nanoribbon terminations: (i) the semiconductor-metallic transition due to the interconvertibility of the Lieb and kagome lattices, (ii) the effect of both nanoribbon width and inclusion of the next-nearest-neighbor hopping term on the degeneracy of the quasi-flat states, (iii) the behavior of the energy gap versus the nanoribbon width, (iv) the existence and evolution of edge states, and (v) the nodal spatial distributions of the total probability densities of the non-dispersive states.  
\end{abstract}

\maketitle

\section{Introduction}\label{sec.introduction}

The isolation of graphene in 2004 \cite{novoselov2004electric} has spurred a search for new two-dimensional (2D) materials aimed at different technological applications.\cite{shim2017electronic, miro2014atlas, khan2020recent} It is known that the type of elements, the hybridization, and the geometry of the formed lattice are key factors for the resulting electronic band structures.\cite{liu2012strategies, li2018graphene, castroNeto} This led to investigations into possible designed crystals with desired geometries and interesting physical properties. Examples of 2D materials engineering are the Lieb \cite{lieb1989,freeney2022electronic} and kagome lattices,\cite{Mielke_1992,neupert2022charge} in which the band structures are formed by the coexistence of a Dirac-like energy band and a flat (non-dispersive) band. Such lattice configurations have motivated research on analogous electronic lattices,\cite{marlou2017, Lieaau4511} waveguide-based photonic systems,\cite{PhysRevLett.114.245503, shen2010, vicencio2015} and even organic structures with the periodicity of these lattices.\cite{2015Natur, Reichardt2018, exp_kagome1}

Interestingly, in 2019, it was realized that Lieb and kagome lattices are interconvertible by applying strains along the diagonal direction, as they share the same structural configuration in the unit cell, \textit{i.e.} one corner-site B and two edge-center sites A and C. \cite{tony2019} This interconvertibility allows the construction of a tight-binding Hamiltonian for a generic lattice dependent on an angle $\theta$ in the range of [$\pi/2$, $2\pi/3$], which describes the Lieb and kagome lattices as the two limiting cases, respectively.\cite{tony2019,lima2022,lang2023tilted} Such angle $\theta$ is defined between the nearest-neighbor (NN) bounds $A$--$B$ and $C$--$B$, as shown in Fig.~\ref{Fig1}(a).  Furthermore, this approach allows studying the transition lattices defined by $\pi/2<\theta<2\pi/3$. 

Inspired by studies that explore finite size effects on the electronic spectrum of one-dimensional nanostructures of layered 2D crystals, such as graphene,\cite{brey2006electronic, wakabayashi2010electronic,son2006energy,way2022graphene} phosphorene,\cite{carvalho2014phosphorene, watts2019production, wu2015electronic, li2022crossed, de2016boundary} silicene,\cite{de2010evidence, de2012multilayer, song2010effects, diniz2022band} transition metal dichalcogenides \cite{aljarb2020ledge, cui2017contrasting, dias2018band, devi2021novel} and $\alpha-\tau_3$ \cite{oriekhov2018electronic, chen2019enhanced, tan2020valley, iurov2021tailoring, hao2022zigzag} lattices, here we investigate the energy spectrum of nanoribbons made out of monolayer Lieb, transition, and kagome lattices using a tight-binding model with a general Hamiltonian that, as mentioned before, takes into account the interconvertibility feature of these lattices.\cite{tony2019} 

Although there are already works in the literature investigating separately finite effects on the electronic properties of Lieb \cite{weeks2010, goldman2011topological, Bandres2014LiebPT, palumbo2015two, van2019interplay, wang2016topological} and kagome \cite{wang2008edge, guo2009, wang2010quantum, liu2010simulating, dey2011magnetic, nadeem2020quantum, zhang2019staggered, mei2020spin, liu2020strain, scammell2022chiral} nanoribbons, none of them, to our knowledge, present a systematic study of the transition lattices in view of the interconvertibility aspect of Lieb-kagome lattices. For instance, Refs.~[\onlinecite{zhang2016dispersion}] and [\onlinecite{chen2016finite}] investigated Lieb nanoribbons with and without intrinsic spin-orbit (ISO) coupling, respectively, for three kinds of edges: straight, bearded, and asymmetric edges [see Figs.~\ref{Fig1}(b), \ref{Fig1}(c), \ref{Fig1}(d)]. The former\cite{zhang2016dispersion} explored dispersion relations of strained complex Lieb lattices based on the tight-binding method without ISO coupling by adopting a NN sites approximation, whereas the latter obtained the energy spectra and wave functions of the edge states analytically for Lieb nanoribbons, the variation of the energy gap for different ribbon widths, and demonstrated that the edge modes could be controlled by applying an on-site potential to the outermost atoms at the boundaries. Moreover, Ref.~[\onlinecite{bo2019exotic}] studied the interplay between the ISO coupling, the lattice modulation, and the magnetic field in Lieb nanoribbons. Concerning the topologically non-trivial edge states, characteristic of a $\mathbb{Z}_2$ classification, it was only identified when ISO coupling is considered.\cite{chen2017spin, bolens2019topological, titvinidze2021, kane2005z, kanemele2005, fu2007topological, qi2008topological} 

Recently, two works\cite{PhysRevA_99_033821, springer2020topological} carried out theoretical investigations on the energy spectra of both Lieb and kagome nanoribbons using tight-binding models which neglect the Lieb-kagome interconvertibility feature, \textit{i.e.} within a disconnected approach for Lieb and kagome lattices. Reference~[\onlinecite{PhysRevA_99_033821}] showed the evolution of Floquet bands of deep longitudinally driven Lieb and kagome waveguide lattices, while Ref.~[\onlinecite{springer2020topological}] compared the effects of including next-nearest-neighbor (NNN)-sites and ISO coupling on the electronic properties of nanoribbons of Lieb, kagome, and other topological 2D polymers. In Ref.~[\onlinecite{tony2019}], Jiang \textit{et al.} reported the first study of both Lieb and kagome nanoribbons within the interconvertibility approach, allowing to explore transition ribbon stages naturally. However, the authors only presented the evolution of the energy spectrum for one type of edge, focusing on the edge states that arise due to the inclusion of ISO coupling. As far as we know, a systematic study of the evolution of the electronic states for transition Lieb-kagome nanoribbons with different terminations has not been carried out, and no previous work has reported edge states for bearded-edged kagome nanoribbons without ISO coupling. These issues will be addressed in the present work. Regarding the edge states, Ref.~[\onlinecite{chen2016finite}] showed that the energy spectra of Lieb nanoribbons do not present edge states in the absence of ISO coupling, whereas Ref.~[\onlinecite{wang2008edge}] demonstrated the existence of such edge states in the case of kagome nanoribbons with asymmetric edges. Here, we shall determine which nanoribbon types support their existence and discuss their degeneracy. 

The paper is organized as follows. In Sec.~\ref{sec.II}, we present the crystallographic aspects of the generic Lieb-kagome nanoribbons (Sec.~\ref{sec.II.A}) and the generic tight-binding Hamiltonian (Sec.~\ref{sec.II.B}), which describes all three Lieb, transition, and kagome nanoribbons by assuming three types of ribbon edges: straight [Fig.~\ref{Fig1}(b)], bearded [Fig.~\ref{Fig1}(c)], and asymmetric [Fig.~\ref{Fig1}(d)]. Results for the energy spectra and probability densities of the bulk, non-dispersive, and edge states are discussed in Sec.~\ref{sec.III}: for different ribbon terminations and widths in Sec.~\ref{Sec.IIIa}, and considering or not NNN interactions in Sec.~\ref{Sec.NNN_sites}. The scaling laws obeyed by the energy band gap for Lieb, transition, and kagome nanoribbons are depicted in Sec.~\ref{Sec.laws_gap}. Finally, in Sec.~\ref{sec.IV} we summarize our main findings.

\section{Theoretical model}\label{sec.II}

\subsection{Crystalographic lattice structure for generic Lieb-kagome nanoribbons}\label{sec.II.A}

\begin{figure}[t]
	\centering
    {\includegraphics[width=\linewidth]{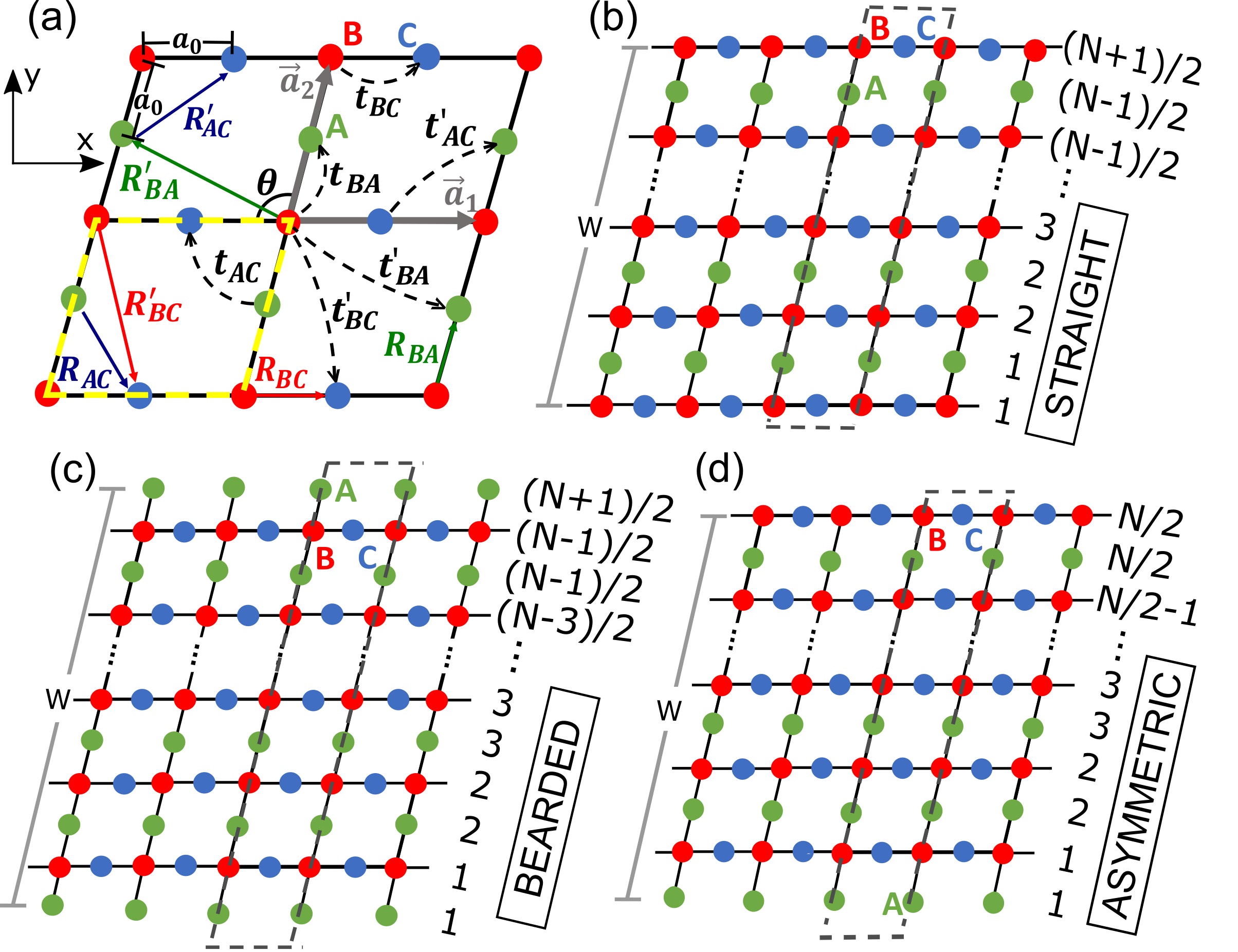}}
	\caption{(Color online) (a) Schematic representation of a generic lattice controlled by the $\theta$-parameter, corresponding to Lieb lattice for $\theta=\pi/2$, transition lattice for $\pi/2<\theta<2\pi/3$, and kagome lattice for $\theta=2\pi/3$. The gray arrows are the primitive vectors $\vec{a}_1$ and $\vec{a}_2$. The unit cell is denoted by a yellow dashed line that contains three non-equivalent sites: A (green), B (red), and C (blue). The distance between NN sites is $a_0$ and the non-null hopping parameters are represented by $t_{BA}$, $t_{BA}^\prime$, $t_{BC}$, $t_{BC}^\prime$, $t_{AC}$, $t_{AC}^\prime$, corresponding to the sites located by $\vec{R}_{BA}=\vec{a}_2/2$, $\vec{{R}}^\prime_{BA}=\left(\vec{a}_2/2-\vec{a}_1\right)$, $\vec{R}_{BC}=\vec{a}_1/2$, $\vec{R}^\prime_{BC}=\left(\vec{a}_1/2-\vec{a}_2\right)$, $\vec{R}_{AC}=\left(\vec{a}_1-\vec{a}_2\right)/2$, and $\vec{R}^\prime_{AC}=\left(\vec{a}_1+\vec{a}_2\right)/2$, respectively. Sketches of generic nanoribbons with (b) straight, (c) bearded, and (d) asymmetric edges, emphasizing their unit cells (gray dashed lines), edge terminations, and the row count in each case are shown. $N$ is the total number of lines defining the ribbon size $W=(N-1)a_0$, which is always odd for straight or bearded edges and even for asymmetric edges.}
	\label{Fig1}
\end{figure}

We consider a 2D generic lattice formed by three base sites [$A$, $B$, and $C$, as indicated by green, red, and blue circles, respectively, in Fig.~\ref{Fig1}(a)]. The nomenclature ``generic lattice'' is due to the $\theta$ control lattice parameter, which allows us to treat, in a general way, both Lieb ($\theta=\pi/2$), transition ($\pi/2<\theta<3\pi/2$), and kagome ($\theta=3\pi/2$) lattices, \textit{i.e.} the present model allows us to map the evolutionary stages of the interconvertibility process between Lieb and kagome lattices through only one parameter.\cite{tony2019, lima2022} The $\theta$ angle is the largest angle between the NN-sites links connecting the sublattices $B-A$ and $B-C$, as shown in Fig.~\ref{Fig1}(a). The primitive vectors, illustrated by gray arrows in Fig.~\ref{Fig1}(a), are given by $\vec{a}_{1}=a(1,0)$ and $\vec{a}_{2}=a(-\cos{\theta},\sin{\theta})$.   

Within this interconvertibility approach, we shall consider Lieb, transition, and kagome nanoribbons with three types of edges: (i) straight [Fig.~\ref{Fig1}(b)], (ii) bearded [Fig.~\ref{Fig1}(c)] and (iii) asymmetric [Fig.~\ref{Fig1}(d)] ribbon terminations. In Figs.~\ref{Fig1}(b), \ref{Fig1}(c), and \ref{Fig1}(d), we show how the ribbon lines are numbered along the finite size $y$ direction, the structural termination edges, and the unit cell (dashed gray line) for the three different investigated nanoribbons. Here, the straight edge [Fig.~\ref{Fig1}(b)] is characterized by its first and last lines being formed by $B$ and $C$ sites. It contains $N_{A}=(N-1)/2$ and $N_{B}=N_{C}=(N+1)/2$ lines of sites $A$, $B$, and $C$, respectively. For the bearded edge case [Fig.~\ref{Fig1}(c)], $A$ type sites form the first and last ribbon lines, whereas for the asymmetric edge case [Fig.~\ref{Fig1}(d)], their terminations are composed by the two previous edges, \textit{i.e.} a straight edge on one side of the nanoribbon and a bearded edge on the other side. In this way, the first line of atoms of the asymmetric nanoribbon is formed by $A$ sublattice type, and its last line by $B$ and $C$ sublattice types. Thus, according to the numbering of the ribbon lines, one has for the bearded edge case $N_{A} =(N+1)/2$ and $N_{B}=N_{C}=(N-1)/2$ lines of sites $A$, $B$, and $C$, respectively, while the asymmetric edge case, one has $N_{A}=N_{B}=N_{C}=N/2$ lines. It is worth mentioning that $N$, the total number of lines defining the ribbon width $W=(N-1)a_0$, is always by our counting way odd for straight or bearded edges, whereas $N$ is necessarily even for asymmetric edges.

\subsection{Tight-binding model for Lieb-kagome nanoribbons}\label{sec.II.B}

Let us now write down the tight-binding Hamiltonian that can describe the charge carrier's dynamics as 
\begin{equation}
    H = \sum_{i,j}t_{i,j}d_{i }^{\dagger}d_{j } + \textrm{H.c.},
\label{Eq.Hamiltonian}
\end{equation}
where $d_{i}^{\dagger}$ ($d_{i}$) creates (annihilates) an electron in the site $i$ of the sublattice $d=\{a,b,c\}$. $t_{i,j}$ are the hopping parameters between sites $i$ and $j$ with displacement vectors given by $\vec{R}_{BA}=\vec{a_2}/2$, $\vec{{R}}^\prime_{BA}=\left(\vec{a}_2/2-\vec{a}_1\right)$, $\vec{R}_{BC}=\vec{a}_1/2$, $\vec{R}^\prime_{BC}=\left(\vec{a}_1/2-\vec{a}_2\right)$, $\vec{R}_{AC}=\left(\vec{a}_1-\vec{a}_2\right)/2$, and $\vec{R}^\prime_{AC}=\left(\vec{a}_1+\vec{a}_2\right)/2$, as represented in Fig.~\ref{Fig1}(a). The corresponding non-null hopping parameters (${t}_{BA}$, ${t}_{BA}^\prime$, ${t}_{BC}$, ${t}_{BC}^\prime$, ${{t}_{AC}}$, ${t}_{AC}^{\prime}$) obey the following equation \cite{lima2022}
\begin{equation}\label{hopping}
	t_{i,j}=te^{-n{\left(a_{ij}/a_0-1\right)}}{a_0}/{a_{ij}}, \qquad n=8,
\end{equation}
where $a_{ij}$ is the distance between the sites $i$ and $j$ and $t$ is the value of the hopping parameter between the NN sites in the generic lattice, \textit{i.e.}~with distance $a_0$ [see Fig.~\ref{Fig1}(a)]. The prime notation used here in the hoppings and displacement vectors denotes connections beyond NNs. Equation~\eqref{hopping}, taking $n=8$, governs the hopping parameters for distances beyond the NN sites in such a way that it leads to an effective tight-binding model for NN sites in the specific cases of Lieb and kagome nanoribbons. However, for transitions between these two lattices, higher-order hopping terms beyond the first NNs become relevant. Thus, it is crucial to investigate the effects of including second and third neighbors, which can be achieved by considering values of $n<8$.\cite{tony2019,lima2022} This approach provides a more comprehensive analysis of the system properties and offers insights into the role of higher-order hopping terms in the system's evolution since the hoppings are functions of the interatomic distances between different sites of the lattice. Reference~[\onlinecite{tony2019}] showed that Eq.~\eqref{hopping} for $n=8$ results in the expected energy bands of the Lieb and kagome lattices, characterized by flat and conic bands. Further details can be found in Appendix~\ref{sec.appendix_A} for the role of $n$ in the dispersion relations of the Lieb-kagome nanoribbons. 

Due to the periodicity of the ribbon structures along the $x$ direction, which are characterized by the number of lines as shown in Figs.~\ref{Fig1}(b)-\ref{Fig1}(d), it is convenient to write our operators in the Fourier basis as
\begin{subequations} \label{eq.d}
\begin{align}
d_{i} &= \frac{1}{\sqrt{N_{cells}}} \sum_{k_{x}} \sum_{n} e^{ik_{x}x_{i}} d_{k_{x},n},\\
d_{j}^{\dagger} &= \frac{1}{\sqrt{N_{cells}}} \sum_{k_{x}'}\sum_{n'} e^{-ik_{x}'x_{j}} d_{k_{x}',n'}^{\dagger},
\end{align}
\end{subequations}
where $N_{cells}$ is the number of unit cells and $d_{k_{x},n}(d_{k_{x}',n'}^{\dagger})$ destroys (creates) an electron with momentum $\hbar k_{x}(\hbar k_{x }')$ at the site of type $d=\{a,b,c\}$ at line $n(n')$. 

Replacing the operators of Eq.~\eqref{eq.d} in Eq.~\eqref{Eq.Hamiltonian}, multiplying the resulting expression by the factor $e^{-ik_{x}'x_{i}} e^{ik_{x}'x_{i}} = 1$, and using the appropriate Kronecker delta function representation, we obtain the following Hamiltonian for generic nanoribbons:
\begin{align}
H & = \sum_{k_{x}} \hspace{-0.05cm}\sum_{n,n'}  \tau_{n',n}^{BA}  a_{k_{x},n}^{\dagger} b_{k_{x},n'} + \sum_{k_{x}'} \hspace{-0.1cm}\sum_{n',n''} \hspace{-0.15cm}\tau_{n',n''}^{BC}  b_{k_{x}',n'}^{\dagger} c_{k_{x}',n''}\hspace{-0.1cm}
\nonumber\\ 
&+ \sum_{k_{x}} \hspace{-0.1cm}\sum_{n,n''} \tau_{n,n''}^{AC}  a_{k_{x},n}^{\dagger} c_{k_{x},n''}\hspace{-0.05cm}
+h.c.,
\end{align}
where $\tau_{n,n'}^{ij}$, with $i,j=(A,B,C)$, are matrix elements defined for each type of nanoribbon termination adopted. Writing $\vec{K}=(k_x,0)$, one can find for straight nanoribbons [Fig.~\ref{Fig1}(b)] that
\begin{subequations}
\begin{align}
    \tau_{n',n}^{BA}&=\delta_{n,n'} (t_{BA} e^{-i\vec{R}_{BA}\cdot\vec{K}} + t_{BA}' e^{-i\vec{R}_{BA}'\cdot\vec{K}})\nonumber\\
	&+ \delta_{n,n'-1} (t_{BA} e^{i\vec{R}_{BA}\cdot\vec{K}} + t_{BA}' e^{i\vec{R}_{BA}'\cdot\vec{K}}),
 \\
    \tau_{n,n''}^{AC} &= \delta_{n,n''} (t_{AC} e^{i\vec{R}_{AC}\cdot\vec{K}} + t_{AC}' e^{-i\vec{R}_{AC}'\cdot\vec{K}})\nonumber\\
	&+ \delta_{n,n''-1} (t_{AC} e^{-i\vec{R}_{AC}\cdot\vec{K}} + t_{AC}' e^{i\vec{R}_{AC}'\cdot\vec{K}}),
 \\
    \tau_{n',n''}^{BC} &= \delta_{n',n''} (t_{BC} e^{i\vec{R}_{BC}\cdot\vec{K}} + t_{BC} e^{-i\vec{R}_{BC}\cdot\vec{K}})\nonumber\\	&+\hspace{-0.05cm}\delta_{n'\hspace{-0.05cm},n''-1} t_{BC}' e^{-i\vec{R}_{BC}'\cdot\vec{K}} \hspace{-0.15cm}+\hspace{-0.05cm} \delta_{n'\hspace{-0.05cm},n''+1} t_{BC}' e^{i\vec{R}_{BC}'\cdot\vec{K}}\hspace{-0.1cm},
\end{align}
\end{subequations}
and for bearded [Fig.~\ref{Fig1}(c)] and asymmetric [Fig.~\ref{Fig1}(d)] nanoribbons, one gets
\begin{subequations}
\begin{align}
    \tau_{n',n}^{BA} &= \delta_{n,n'} (t_{BA} e^{i\vec{R}_{BA}\cdot\vec{K}} + t_{BA}' e^{i\vec{R}_{BA}'\cdot\vec{K}})\nonumber\\
	&+ \delta_{n,n'+1} (t_{BA} e^{-i\vec{R}_{BA}\cdot\vec{K}} + t_{BA}' e^{-i\vec{R}_{BA}'\cdot\vec{K}}),
 \\
    \tau_{n,n''}^{AC} &= \delta_{n,n''} (t_{AC} e^{-i\vec{R}_{AC}\cdot\vec{K}} + t_{AC}' e^{i\vec{R}_{AC}'\cdot\vec{K}})\nonumber\\
	&+ \delta_{n,n''+1} (t_{AC} e^{i\vec{R}_{AC}\cdot\vec{K}} + t_{AC}' e^{-i\vec{R}_{AC}'\cdot\vec{K}}),
 \\
    \tau_{n',n''}^{BC} &= \delta_{n',n''} (t_{BC} e^{i\vec{R}_{BC}\cdot\vec{K}} + t_{BC} e^{-i\vec{R}_{BC}\cdot\vec{K}})\nonumber\\	&+\hspace{-0.05cm}\delta_{n'\hspace{-0.05cm},n''-1} t_{BC}' e^{-i\vec{R}_{BC}'\cdot\vec{K}} \hspace{-0.15cm}+\hspace{-0.05cm} \delta_{n',n''+1} t_{BC}' e^{i\vec{R}_{BC}'\cdot\vec{K}}\hspace{-0.15cm}.
\end{align}
\end{subequations}

To find the Lieb-kagome nanoribbon band structure, we apply the standard Heisenberg equation of motion, $i\hbar d \mathcal{O}/dt=[\mathcal{O},H]$, to the operators $\mathcal{O} \equiv \{d_{k_x,n}^\dagger$ and $d_{k_x,n}\}$ in line $n$. Assuming that the time dependence of the modes behaves like $e^{-\frac{iEt}{\hbar}}$, one obtains $Ed_{k_{x},n}=[d_{k_{x},n}, H]$ for the three operators $d\equiv \{a,b,c\}$, and thus  
\begin{subequations}
    \begin{align}
&[a_{k_{x},n},H]= \sum_{n'=1}^{N_{B}} \tau_{n,n'}^{AB} b_{k_{x},n'}+\sum_{n'=1}^{N_{C}} \tau_{n,n'}^{AC} c_{k_{x},n'},
\label{C1}
\\
&
[b_{k_{x},n},H]= \sum_{n'=1}^{N_{A}} \tau_{n,n'}^{BA} a_{k_{x},n'} + \sum_{n'=1}^{N_{C}} \tau_{n,n'}^{BC} c_{k_{x},n'},
\label{C2}
\\
&[c_{k_{x},n},H]= \sum_{n'=1}^{N_{B}} \tau_{n,n'}^{BC} b_{k_{x},n'}+\sum_{n'=1}^{N_{A}} \tau_{n,n'}^{AC} a_{k_{x},n'}.
\label{C3}
    \end{align}
\end{subequations}
Combining this set of coupled equations [Eqs.~\eqref{C1}, \eqref{C2}, and \eqref{C3}], one arrives at the following matrix equation
\begin{equation}
\mathcal{T} \begin{pmatrix}
                 a_{k_{x},n} \\
                 b_{k_{x},n} \\
                 c_{k_{x},n}
\end{pmatrix} = E \begin{pmatrix}
                  a_{k_{x},n} \\
                  b_{k_{x},n} \\
                  c_{k_{x},n}
\end{pmatrix},
\end{equation}
where $\mathcal{T}$ is the hopping matrix of order $N_{A}+N_{B}+N_{C}$ that depends on the
ribbon configuration, given by
\begin{equation}\label{eq.matrix_T}
\mathcal{T} \hspace{-0.05cm}=\hspace{-0.05cm}\begin{pmatrix}
       [\tau^{AA}]_{N_{A}\times N_{A}} & [\tau^{AB}]_{N_{A}\times N_{B}} & [\tau^{AC}]_{N_{A}\times N_{C}} \\
        [\tau^{BA}]_{N_{B}\times N_{A}} & [\tau^{BB}]_{N_{B}\times N_{B}} &  [\tau^{BC}]_{N_{B}\times N_{C}} \\
       [\tau^{CA}]_{N_{C}\times N_{A}} &  [\tau^{CB}]_{N_{C}\times N_{B}} & [\tau^{CC}]_{N_{C}\times N_{C}}
\end{pmatrix},
\end{equation}
with $\tau^{AA}$, $\tau^{BB}$, and $\tau^{CC}$ standing for the on-site energies of sites $A$, $B$, and $C$, respectively, that are assumed here to be zero. To illustrate an example of a non-null on-site energy case, one has that Lieb-kagome nanoribbons subjected to perpendicular or in-plane electric fields are simulated by changing the on-site energies in the corresponding appropriate way. As discussed in Sec.~\ref{sec.II.A}, according to the number of lines for each Lieb-kagome nanoribbon, the order of the matrix $\mathcal{T}$ is $(3N-1)/2$, $(3N+1)/2$, and $3N/2$ for nanoribbons with bearded, straight, and asymmetric edges, respectively. Therefore, straight-edged nanoribbons have an additional mode compared to bearded-edged nanoribbons of the same size, as shall be discussed in Sec.~\ref{sec.III}. 

To obtain the energy levels, the hopping matrix $\mathcal{T}$ is diagonalized by choosing the ribbon features, such as the angle $\theta$, that determines the lattice type, the type of edge, and the total number of lines $N$ associated with the width of the nanoribbon. With the energies, one can calculate the density of states, which describes the proportion of states to be occupied at each certain energy range, \textit{i.e.}~its evaluation is performed by a superposition of individual energy states that we broaden using a Gaussian function\cite{da2015energy} 
\begin{equation}
f(E)=e^{-(E-E_0)^2/\gamma^2},
\end{equation}
with a broadening factor $\gamma$ smaller than the energy level separations. Unless otherwise stated, we assume $\gamma=0.05t$ for all density of states results from here onward.

\section{Results and discussions}\label{sec.III}

\subsection{Straight, bearded and asymmetric nanoribbons}\label{Sec.IIIa}

\begin{figure*}[htbp]
	\centering
{\includegraphics[width=\linewidth]{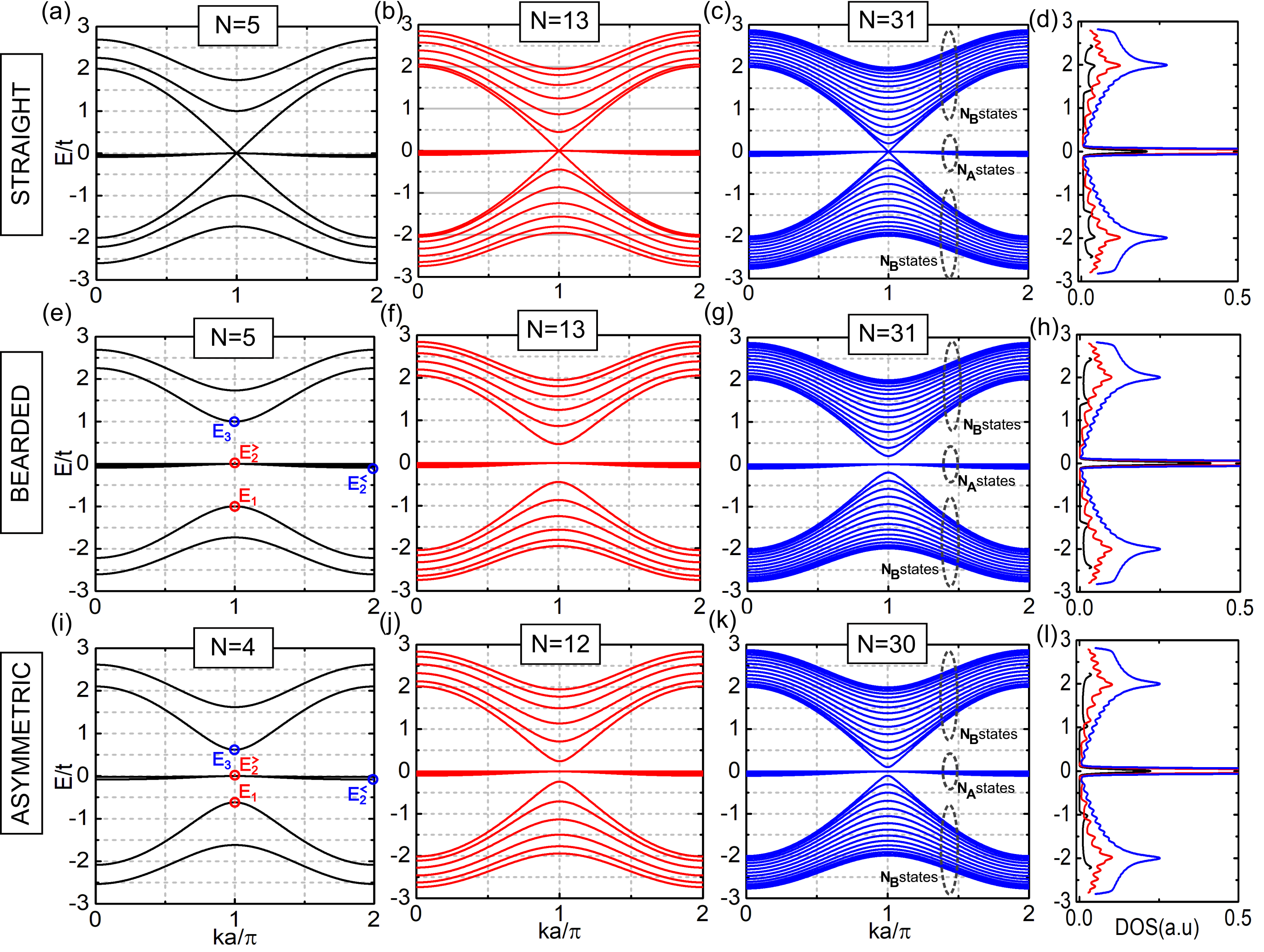}}
\caption{(Color online) Band structures and density of states for Lieb nanoribbons, \textit{i.e.}~$\theta = 90^\circ$, with (top row) straight, (middle row) bearded, and (bottom row) asymmetric edges, and different numbers of atomic lines. The density of states' panels are depicted by black, red, and blue curves, corresponding to the energy spectra of the first, second, and third-panel columns, respectively. $E_1$ corresponds to the higher energy of the lower dispersive subband, $E_2^<$ ($E_2^>$) the lower (higher) energy of the quasi-flat subband, and $E_3$ the lower energy of the upper dispersive subband. Non-dispersive and bulk states are grouped with $N_A$ and $N_B$ levels, respectively, as indicated in panels (c), (g), and (k).}
\label{Fig2}
\end{figure*}

Figure~\ref{Fig2} depicts Lieb nanoribbons' energy levels for the three investigated terminations: (i) straight edges with $N=5$ [Fig.~\ref{Fig2}(a)], $N=13$ [Fig.~\ref{Fig2}(b)], and $N=31$ [Fig.~\ref{Fig2}(c)]; (ii) bearded edges with $N=5$ [Fig.~\ref{Fig2}(e)], $N=13$ [Fig.~\ref{Fig2}(f)], and $N=31$ [Fig.~\ref{Fig2}(g)]; (iii) asymmetric edges with $N=4$ [Fig.~\ref{Fig2}(i)], $N=12$ [Fig.~\ref{Fig2}(j)], and $N=30$ [Fig.~\ref{Fig2}(k)]. The corresponding density of states for these three terminations are shown in Figs.~\ref{Fig2}(d), \ref{Fig2}(h), and \ref{Fig2}(l), respectively. Such results reveal the presence of $N_A$ quasi-flat states localized around $E=0$ and $N_B$ energy levels for each of both the lower (negative) and the upper (positive) subbands, originating from the infinite-sheet flat and dispersive bands, respectively, as indicated in Figs.~\ref{Fig2}(c), \ref{Fig2}(g), and \ref{Fig2}(k). Notably, the Dirac cone observed in the energy spectrum of the infinite-sheet Lieb lattice\cite{tony2019,lima2022} is verified only in the Lieb ribbon case with straight edges [Figs.~\ref{Fig2}(a), \ref{Fig2}(b), \ref{Fig2}(c)]. This can be linked to the fact that straight-edged nanoribbon is the only one of the three edge terminations that is completely symmetrical and with no dangling bonds, thus preserving the defect-free structural Lieb lattice features and consequently exhibiting the main energetic Lieb aspects in its infinite-sheet spectrum, \textit{i.e.}~the coexistence of a Dirac cone and quasi-flat band. For Lieb nanoribbons with bearded edges [Figs.~\ref{Fig2}(e), \ref{Fig2}(f), \ref{Fig2}(g)] and asymmetric edges [Figs.~\ref{Fig2}(i), \ref{Fig2}(j), \ref{Fig2}(k)], one can notice the appearance of two energy gaps: one of these gaps situated between the higher energy ($E_1$) of the lower dispersive subband and the lower energy ($E_2^<$) of the quasi-flat subband, and the other energy gap emerging between the higher energy ($E_2^>$) of the quasi-flat subband and the lower energy ($E_3$) of the upper dispersive subband, as indicated in Figs.~\ref{Fig2}(e) and \ref{Fig2}(i). \cite{zhang2016dispersion} As shall be discussed in detail in Sec.~\ref{Sec.laws_gap}, the size of these gaps are different when comparing bearded and asymmetric Lieb nanoribbons, with the bearded edge case exhibiting a wider gap, \cite{zhang2016dispersion} but their scaling laws showing a similar tendency. Based on the presence or absence of an energy gap in the energy spectrum, one can characterize straight-edged Lieb nanoribbons as metallic and bearded and asymmetric-edged Lieb nanoribbons as semiconductor systems. 

The density of states panels [Figs.~\ref{Fig2}(d), \ref{Fig2}(h), and \ref{Fig2}(l)] show a pronounced sharp peak around $E=0$ associated with the $N_A$ quasi-degenerate states and its peak width related to the quasi-flat states subband width. Additionally, around $E/t \approx \pm 2$, the density of states of the Lieb nanoribbons exhibits van Hove singularities (peaks) corresponding to energy levels with almost zero group velocity at the corners of the first Brillouin zone, presenting a higher peak the larger the Lieb nanoribbon (compare black and blue curves, where the former case represents the shortest-sized Lieb nanoribbon). These van Hove singularities were also reported in other flat-band nanostructured systems, such as Dice lattice nanoribbons \cite{soni2020flat}.

\begin{figure*}[t]
\centering
{\includegraphics[width=\linewidth]{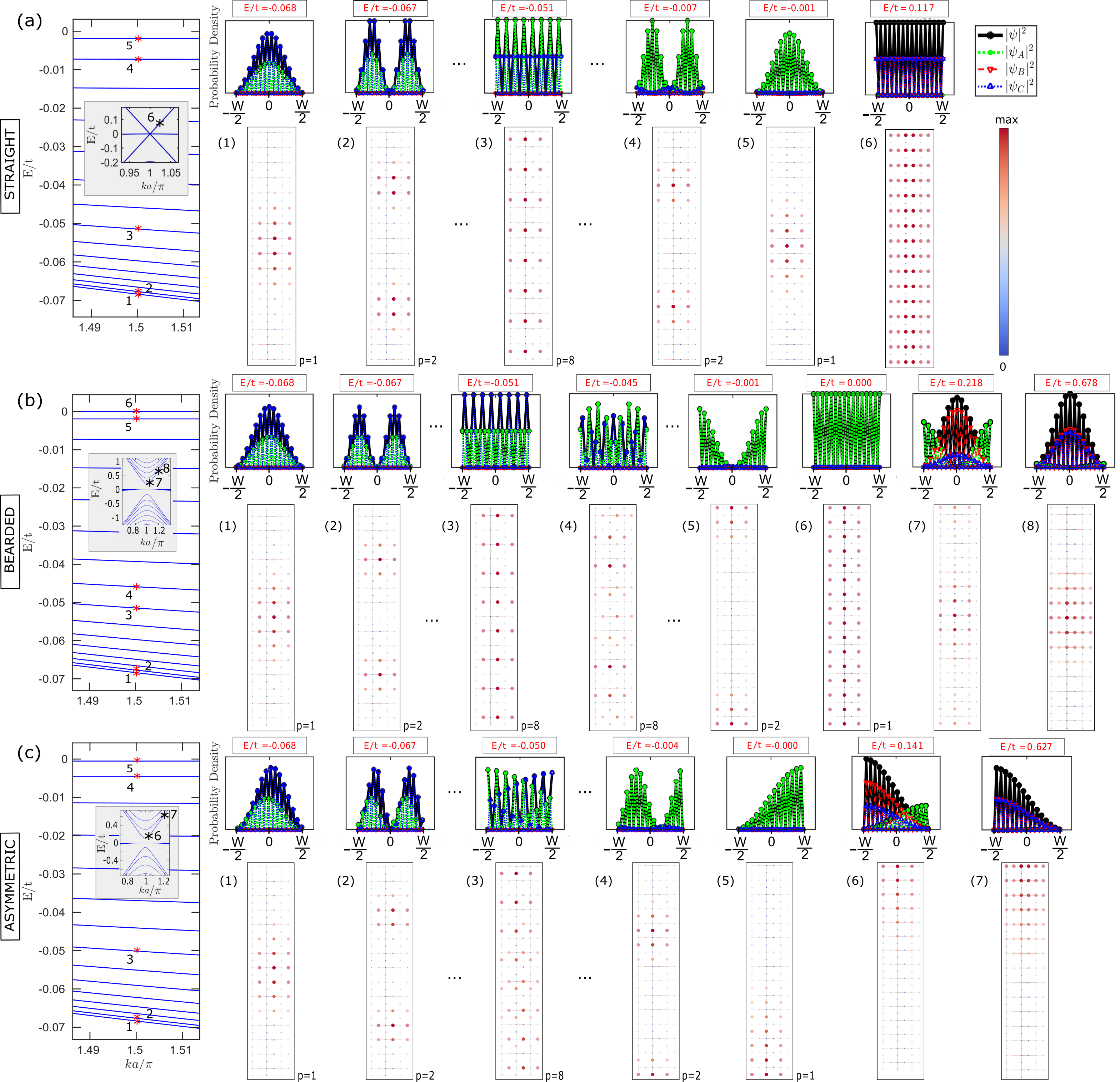}}
\caption{(Color online) 
Probability density distribution of Lieb nanoribbons with (a) straight [$N=31$], (b) bearded [$N=31$], and (c) asymmetric edges [$N=30$], showing (i) the nodal behavior of the electronic states associated with the quasi-flat bands [highlighted by red asterisks in the dispersion relations in (a) (1)--(5), (b) (1)--(6), and (c) (1)--(5), and enumerated from bottom to top at $ka=3\pi/2$], (ii) an equally distributed electronic state in $B-C$ sublattices associated with the Dirac-like band characteristic of the straight edges at $ka=3.261$ [highlighted by a black asterisk (6) in the inset of panel (a)], (iii) bulk states at $ka=\pi+0.1$ (7) [(6)] and $ka=1.2 \pi$ (8) [(7)] highlighted by black asterisks in the inset of panel (b) [(c)]. For each edge type, the top panels show cross-sections of the total probability density [black solid curves with asterisks] and of the modulus squared of the probability amplitudes $|\psi_A|^2$, $|\psi_B|^2$, and $|\psi_C|^2$, related to sublattices A [green short-dashed curves with circles], B [red dashed curves with downward-pointing triangles], and C [blue dotted curves with upward-pointing triangles], respectively. In the lower panels, the circular symbol sizes and the color scale ranging from blue to red represent the amplitude of the probability density ranging from zero to its maximum value along the nanoribbon. The parameter $p$ denotes the number of peaks corresponding to the nodal behavior of the probability densities of the quasi-flat bands.}
\label{Fig3}
\end{figure*}

To understand the nature of the electronic states in the dispersion relations of Fig.~\ref{Fig2}, it is relevant to investigate how the spatial distribution of the wave function is associated with the bulk, quasi-flat, and Dirac-like states. For that, we show in Fig.~\ref{Fig3} cross-section (top panels) and 2D scatter (bottom panels) plots of the total probability densities and the probability amplitudes $|\psi_A|^2$, $|\psi_B|^2$, and $|\psi_C|^2$ related to sublattices A [green short-dashed curves with circles], B [red dashed curves with downward-pointing triangles], and C [blue dotted curves with upward-pointing triangles], respectively, for some specific states marked with asterisks in the dispersion relations in Figs.~\ref{Fig3}(a), \ref{Fig3}(b), and \ref{Fig3}(c) for straight, bearded, and asymmetric Lieb nanoribbons with $N=31$, $N=31$, and $N=30$ atomic lines in the unit cell zoomed from the quasi-flat energy regions of Figs.~\ref{Fig2}(c), \ref{Fig2}(g), and \ref{Fig2}(k), respectively.

Interestingly, the results depicted in Fig.~\ref{Fig3} demonstrate that the quasi-flat states observed in the energy spectra of Lieb nanoribbons with straight [panels (1) to (5) in Fig.~\ref{Fig3}(a)], bearded [panels (1) to (6) in Fig.~\ref{Fig3}(b)], and asymmetric edges [panels (1) to (5) in Fig.~\ref{Fig3}(c)] are not edge states, as may be expected due to its almost non-dispersive character, but rather they are nodal bulk-like states. That is in contrast to the partially flat bands found in zigzag graphene nanoribbons, which correspond to localized edge states at the Fermi energy.\cite{fujita1996peculiar, wakabayashi1999electronic}

From the nodal behavior for the straight-edged Lieb nanoribbon [Fig.~\ref{Fig3}(a)] depicted in panels (1) to (5) for the corresponding states of the quasi-flat subband marked in Fig.~\ref{Fig3}(a), one can notice that the number of peaks in the probability density increases, starting from the lowest almost quasi-flat state [panel (1)], until the $N_A/2$--th state, that in the $N=31$ case corresponds to $8$ as shown in panel (3) and which has eight peaks ($p=8$), and then the number of peaks decreases as the energy value increases until the $N_A$--th state, \textit{i.e.}~$p\in [1,2,\cdots,(N_A-1)/2,N_A/2,(N_A-1)/2,\cdots,2,1]$, with $p$ denoting the number of peaks. Note that the probability densities for the quasi-flat bulk-like states of the straight-edged Lieb nanoribbons are mainly distributed in the C and A sites, as can be clearly seen in the top cross-section panels (1) to (5) by the non-null intensities for $|\psi_A|^2$ in short-dashed green curve with green circles and for $|\psi_A|^2$ in dotted blue curve with blue up-triangles, whereas the $|\psi_B|^2$ in dashed red curve with down-triangles exhibit a null contribution. Additionally, one can realize that for $p\in [1,2,\cdots,(N_A-1)/2]$, the total wave function contributions are more intense on C sites than on A sites, whereas for $p\in [N_A/2,(N_A-1)/2,\cdots,2,1]$ this is the opposite. For instance, by comparing the most [panel (5) in Fig.~\ref{Fig3}(a)] and the lowest [panel (1) in Fig.~\ref{Fig3}(a)] energetic quasi-flat energy levels that have one peak distributed in the central region of the nanoribbon, one observes that the former is more intensively distributed on type A sites and the latter on type C atoms.

The probability densities of the quasi-flat energy levels for Lieb nanoribbons with bearded [Fig.~\ref{Fig3}(b)] and asymmetric [Fig.~\ref{Fig3}(c)] edges also exhibit nodal behavior, similarly to the straight case [Fig.~\ref{Fig3}(a)]. Note that the quasi-flat states in panels (3) and (4) in Fig.~\ref{Fig3}(b) exhibit both $p=8$ peaks. Lieb nanoribbons with bearded edges [$N_A=(N+1)/2$ with $N$ being odd] have one more A sublattice than the straight [$N_A=(N-1)/2$ with $N$ being odd] and asymmetric [$N_A=N/2$ with $N$ being even] cases. Because of that, the nodal peak counting of the quasi-flat states for the bearded-edged case obeys the following sequence $p\in [1,2,\cdots,(N_A-1)/2,N_A/2,N_A/2,(N_A-1)/2,\cdots,2,1]$. In addition, for the bearded edge nanoribbon case, the most energetic state of the quasi-flat subband [panel (6) in Fig.~\ref{Fig3}(b)] does not exhibit a well-defined peak located in the nanoribbon's center, but instead, it shows a constant distribution of electron density at A sites along the whole length of the nanoribbon. To understand that, it is important to keep in mind that both boundaries of the bearded-edged Lieb nanoribbon are formed by dangling bonds and that charge carriers prefer to be localized around them, as happens, for instance, for graphene\cite{song2010dangling, miranda2022vacancy} and AlN\cite{zhang2011effects, sun2014effects} nanoribbons with defective edges. This is the reason why the last half of the energy levels that are most energetic of the quasi-flat subband, corresponding to $p\in [N_A/2,(N_A-1)/2,\cdots,2,1]$, present spatial distributions of their wave functions with pronounced peaks around the edges formed by dangling bonds, as can be viewed in panel (5) in Fig.~\ref{Fig3}(b) with an asymmetric peak in each of the two ribbon's edges ($p=2$). Knowing that and expecting a one-peak nodal-like state for $p=1$ [panel (6) in Fig.~\ref{Fig3}(b)], one ends up, in turn, with the charge carrier wave function being pushed to the nanoribbon's boundaries leading to a uniformly equally distributed probability amplitude. Furthermore, energetically speaking, this last energetic state of the quasi-flat subband is the most non-dispersive mode of all, resulting in a nearly null group velocity and, consequently, a non-preferential localization in a certain region of the nanoribbon. Due to this preferential localization of the wave function around the dangling bonds, the $N_B$ dispersive states for bearded Lieb nanoribbons exhibit a combination of typical nodal behavior of bulk states as expected in quantum confinement systems \cite{harrison2016quantum} with a contribution of the wave function distributions also at the edges. This can be verified in panel (7) at $ka=\pi+0.1$ in Fig.~\ref{Fig3}(b) for the lowest energy level of the upper dispersive subband for bearded Lieb nanoribbon, presenting one peak at the nanoribbon's center with the largest contributions being from the B and C sites and the edge contributions being mainly from A sites. For the same energy level shown in (7) but now with a high momentum value ($ka=1.2\pi$), we show in panel (8) of Fig.~\ref{Fig3}(b) that, indeed, the first conduction energy level is a bulk mode presenting the expected tendency of a well-defined one peak, being less sensitive to edge effects the higher the $ka$-value, as can be seen by the decreasing edge amplitudes from A sites in panel (8) when compared with panel (7). These results of the sublattice contributions associated with the flat and dispersive bands are consistent with the differential conductance maps presented in Ref.~[\onlinecite{marlou2017}], where they provided experimental realizations and characterizations of geometric arrangements of CO molecules on a Cu(111) surface to generate an electronic Lieb lattice.

\begin{figure*}[t]
	\centering   {\includegraphics[width=\linewidth]{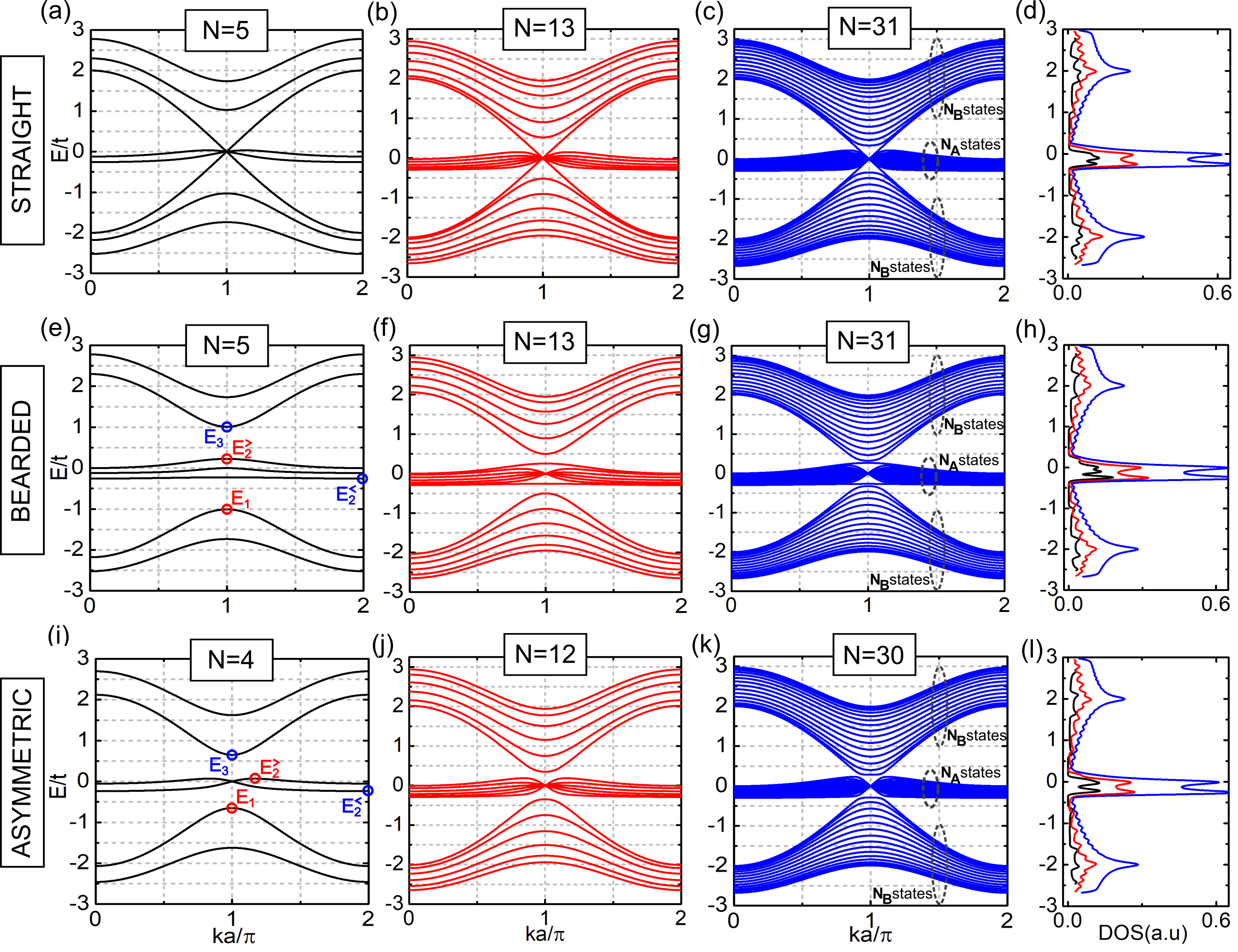}}
	\caption{(Color online) The same as in Fig.~\ref{Fig2}, but now for transition nanoribbon with $\theta = 105^\circ$.}
	\label{Fig4}
\end{figure*}

Similarly to the straight-edged case, for asymmetric-edged Lieb nanoribbons [panels (1) to (5) in Fig.~\ref{Fig3}(c)], the number of peaks of the quasi-flat subband follows the counting $p\in [1,2,\cdots,(N_A-1)/2,N_A/2,(N_A-1)/2,\cdots,2,1]$ with A and C sublattices giving the largest contributions for the total probability densities. Note from panels (1) to (3) in Fig.~\ref{Fig3}(c) that the first half of states $p\in [1,2,\cdots,(N_A-1)/2,N_A/2]$ have their wave function's nodal behavior being mainly formed by the C sublattices and their total probability densities exhibiting nearly symmetric peaks' distributions, whereas the second half of states $p\in [(N_A-1)/2,\cdots,2,1]$ the largest contributions are from A sublattices with their total probability densities being asymmetric. The reason for that is related to the crystallographic structure itself of the asymmetric Lieb nanoribbons that are composed of a straight boundary with edge termination formed here with B and C sublattices and a bearded boundary with edge termination formed here with A sublattices. Thus, for the second half of states in the quasi-flat subband, one gets probability density distributions being pushed to the nanoribbon's boundary with the bearded edge in which the largest sublattice amplitudes are from A sublattices, leading an asymmetric behavior for the spatial distribution, as also observed for the second half of states shown in panels (4), (5) and (6) in Fig.~\ref{Fig3}(b) for the bearded case. It is worth emphasizing that the state (5) in Fig.~\ref{Fig3}(c) is not an edge state, as could perhaps be speculated by the edge localization with an amplitude decay of $|\psi|^2$, but rather, it is a $p=1$ nodal bulk-like state with its preferential localization around the dangling bonds, \cite{song2010dangling, dutta2010novel} exhibiting thus an asymmetric peak pushed to the bearded edge. As a consequence of the two non-equivalent edges that constitute the asymmetric Lieb nanoribbon, the $N_B$ dispersive states also exhibit asymmetrical peaks distributed along the width of the nanoribbon, as for instance, as shown in panels (6) and (7) of Fig.~\ref{Fig3}(c) for the first electron state of the dispersive subbands (one peak - $p=1$) with two different $ka$-values, exhibiting asymmetrical peaks formed around the edges and its maximum peak amplitude being formed mainly for B and C sublattices' contributions the higher the $ka$ value, as verified by comparing panels (6) for $ka=\pi+0.1$ and (7) for $ka=1.2\pi$.

As previously discussed, the emergence of the Dirac-like conic state intersecting the quasi-flat subband in the energy spectrum of the straight-edged Lieb nanoribbons [Figs.~\ref{Fig2}(a), \ref{Fig2}(b), \ref{Fig2}(c), and \ref{Fig3}(a)] is a consequence of the defect-free crystallographic aspect of its edge termination that presents similar energetic features as observed in the infinite-sheet Lieb spectrum. The wave function distribution for this Dirac-like state is shown in panel (6) in Fig.~\ref{Fig3}(a), being a bulk state and exhibiting equally distributed amplitudes on B and C sublattices. Chen \textit{et al.} demonstrated in Ref.~[\onlinecite{chen2016finite}] that such bulk state becomes an edge state when the ISO coupling is taken into account, \textit{i.e.} the ISO coupling induces such Dirac-like bulk states to localize at the edges to become the helical edge states with the same Dirac-like spectrum.

\begin{figure*}[t]
\centering
{\includegraphics[width=\linewidth]{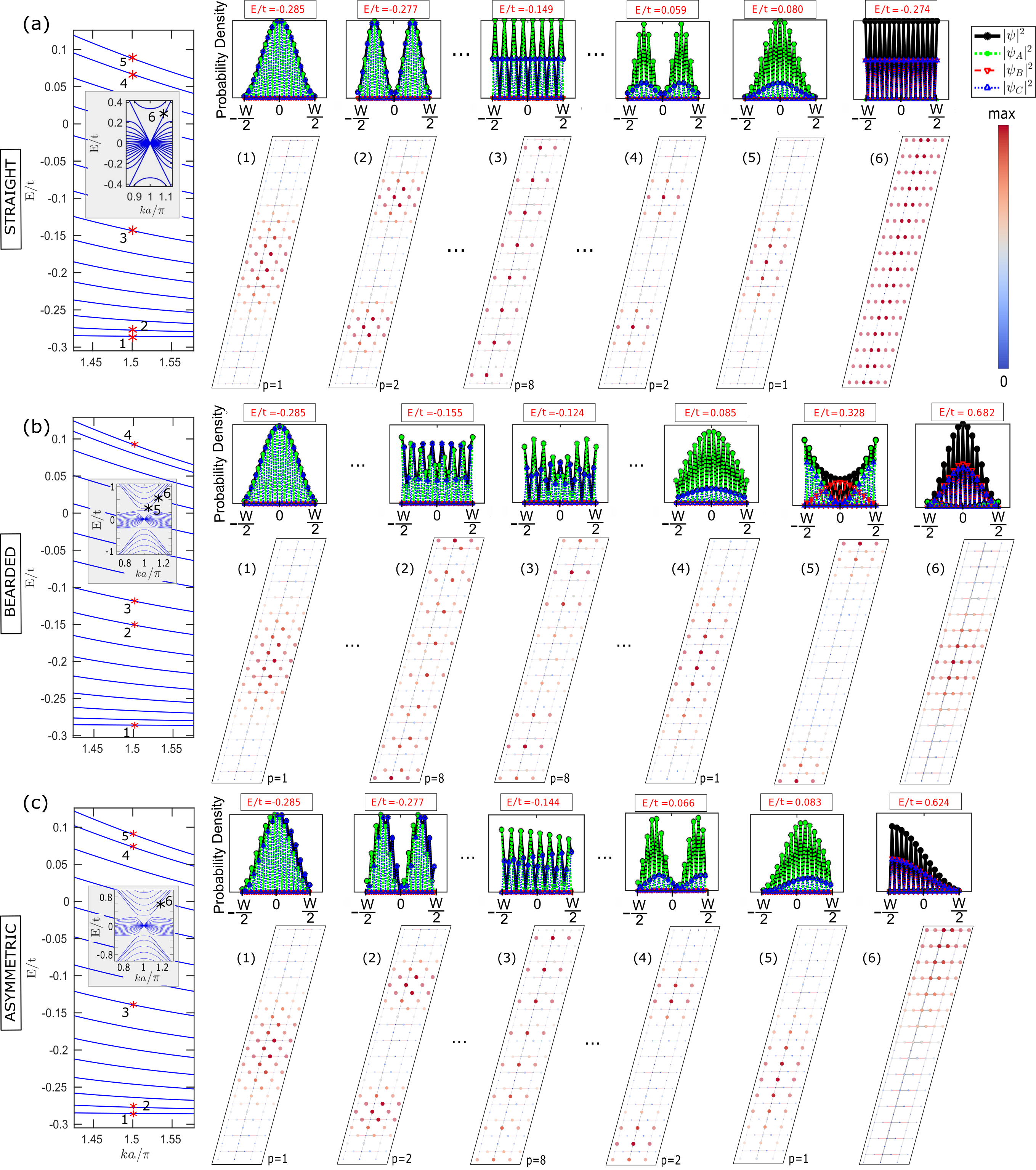}}
\caption{(Color online) The same as in Fig.~\ref{Fig3}, but now for a transition nanoribbon with $\theta=105^\circ$. The quasi-flat bands have also been labeled from bottom to top at $ka=3\pi/2$ [red asterisks], and additionally, (i) the Dirac-like band characteristic of straight edges has been calculated for $ka=3.418$ [see inset in (a) and the black asterisk for the state labeled in (6)], and (ii) the one-peak bulk states for the first energy level of the conduction subband have been shown at $ka=\pi+0.1$ and $ka=1.2 \pi$ [see insets in (b) and (c) and black asterisks for the states labeled in (5 and 6) and (6 and 7)].}
\label{Fig5}
\end{figure*}

Figures~\ref{Fig4} and \ref{Fig5} depict the band structures and probability densities, respectively, for transition nanoribbons assuming an angle of $\theta=105^{\circ}$ in the evolutionary stage in the interconvertibility process of Lieb-Kagome nanoribbons [see Fig.~\ref{Fig1}] and taking the three investigated edge terminations and three different ribbon widths: (i) straight edges with $N=5$ [Fig.~\ref{Fig4}(a)], $N=13$ [Fig.~\ref{Fig4}(b)], and $N=31$ [Fig.~\ref{Fig4}(c)]; (ii) bearded edges with $N=5$ [Fig.~\ref{Fig4}(e)], $N=13$ [Fig.~\ref{Fig4}(f)], and $N=31$ [Fig.~\ref{Fig4}(g)]; (iii) asymmetric edges with $N=4$ [Fig.~\ref{Fig4}(i)], $N=12$ [Fig.~\ref{Fig4}(j)], and $N=30$ [Fig.~\ref{Fig4}(k)]. The corresponding density of states for these three terminations are shown in Figs.~\ref{Fig4}(d), \ref{Fig4}(h), and \ref{Fig4}(l), respectively, and the probability densities for the states labeled with asterisks in the straight-edged [Fig.~\ref{Fig5}(a)], bearded-edged [Fig.~\ref{Fig5}(b)], and asymmetric-edged [Fig.~\ref{Fig5}(c)] transition nanoribbons' dispersion relations are also presented. Generally speaking, by comparing the dispersion relations for Lieb [Fig.~\ref{Fig2}] and transition [Fig.~\ref{Fig4}] nanoribbons, one realizes that the energy spectra of transition nanoribbons differ from the Lieb ones mainly in the shape of the quasi-flat modes, which are more dispersive than those of Lieb nanoribbons, and a larger broadening in the bandwidth of its quasi-flat subband. It can also be verified in the density of states of transition nanoribbons for all three edge configurations [Figs.~\ref{Fig4}(d), \ref{Fig4}(h), and \ref{Fig4}(l)], in which exhibit a splitting of the pronounced and sharp peak characteristic of the flat band at $E=0$ for the Lieb lattice [Figs.~\ref{Fig2}(d), \ref{Fig2}(h), and \ref{Fig2}(l)] into two peaks merged around $E=0$ with a broadened bandwidth. This behavior is a consequence of the interconvertibility process of the transition nanoribbon, in which each one of these two peaks around $E=0$ for transition nanoribbons originates from a pair of titled Dirac cones in the band structure of transition lattice in the first Brillouin zone \cite{lima2022, tony2019}, being one pair of cones formed from the connection between the upper and middle bands (type-I Dirac point), and the other pair of cones from the lower and middle bands (type-II Dirac point). Therefore, this splitting effect provides evidence of the interconvertibility process between Lieb and kagome nanoribbons since the interconvertibility range limit lattices do not present such a characteristic in the density of states. 

Delving deeper into the understanding of transition nanoribbons' spectra [Fig.~\ref{Fig4}], identifying the similarities and differences when compared to Lieb nanoribbons [Fig.~\ref{Fig2}], one can observe that both Lieb [Figs.~\ref{Fig2}(a), \ref{Fig2}(b), and \ref{Fig2}(c)] and transition [Figs.~\ref{Fig4}(a), \ref{Fig4}(b), and \ref{Fig4}(c)] nanoribbons with straight edges are metallic and exhibit the Dirac-like energy state. From panel (6) in Fig.~\ref{Fig5}(a), one notices that this Dirac-like energy level, resembling the Dirac cone present in defect-free Lieb and transition lattices, corresponds to a bulk state with its wave function having equally distributed amplitudes on B and C sublattices along the transition nanoribbon, similarly as observed in panel (6) in Fig.~\ref{Fig3}(a) for Lieb nanoribbon with straight edges. Additionally, the main difference between the energy spectra of transition nanoribbons with bearded [Figs.~\ref{Fig4}(e), \ref{Fig4}(f), and \ref{Fig4}(g)] and asymmetric [Figs.~\ref{Fig4}(i), \ref{Fig4}(j), and \ref{Fig4}(k)] edges is manifested in the quasi-flat bands. In the case of the asymmetric edge, all states of the quasi-flat subband are degenerate at $ka=\pi$--point [see Figs.~\ref{Fig4}(i), \ref{Fig4}(j), and \ref{Fig4}(k)], whereas in the case of the bearded edge, this degeneracy is lifted for the less energetic and the most energetic quasi-flat states in the vicinity of the $ka=\pi$--point [see Figs.~\ref{Fig4}(e), \ref{Fig4}(f), and \ref{Fig4}(g)]. Similarly, as in the case of Lieb nanoribbons [Fig.~\ref{Fig3}], bearded-edged transition nanoribbons have an extra quasi-flat mode when compared to the straight-edged and asymmetric-edged transition nanoribbons, with nodal behavior displaying a peak counting such as $p\in [1,2,\cdots,(N_A-1)/2,N_A/2,N_A/2,(N_A-1)/2,\cdots,2,1]$, as can be seen in panel (1) to (4) in Fig.~\ref{Fig5}(b), in a similar manner as in panels (1) to (6) in Fig.~\ref{Fig3}(b). However, the total probability densities present different sublattice contributions for Lieb [Fig.~\ref{Fig3}] and transition [Fig.~\ref{Fig5}] nanoribbons for the quasi-flat bulk-like states. Although in both cases of Lieb and transition nanoribbons, the quasi-flat states are composed of non-null amplitudes only of $|\psi_A|^2$ and $|\psi_C|^2$ for C and A sublattices, their spatial contributions in the composition of the total wave function are different for Lieb and transition cases. As observed in the top cross-section panels of Fig.~\ref{Fig5} for the first half of quasi-flat states, \textit{i.e.} $p\in [1,2,\cdots,(N_A-1)/2]$ states for straight [see panels (1) to (4) in Fig.~\ref{Fig5}(a)] and asymmetric [see panels (1) to (5) in Fig.~\ref{Fig5}(c)] transition nanoribbons and $p\in [1,2,\cdots,(N_A-1)/2, N_A/2]$ states for bearded-edged transition nanoribbons [see panels (1) to (4) in Fig.~\ref{Fig5}(b)], the total wave function contributions are nearly equally distributed on C and A sites, whereas for the second half of quasi-flat states, \textit{i.e.} $p\in [N_A/2,(N_A-1)/2,\cdots,2,1]$ states, the total wave function contributions are more intense on A sites than on C sites. This is not the scenario observed in Fig.~\ref{Fig3} for the Lieb nanoribbon case, in which no quasi-flat state has its wave function distribution composed of nearly equally distributed amplitudes for A and C sublattices for any of the three edge termination types.

\begin{figure*}[t]
	\centering
  {\includegraphics[width=\linewidth]{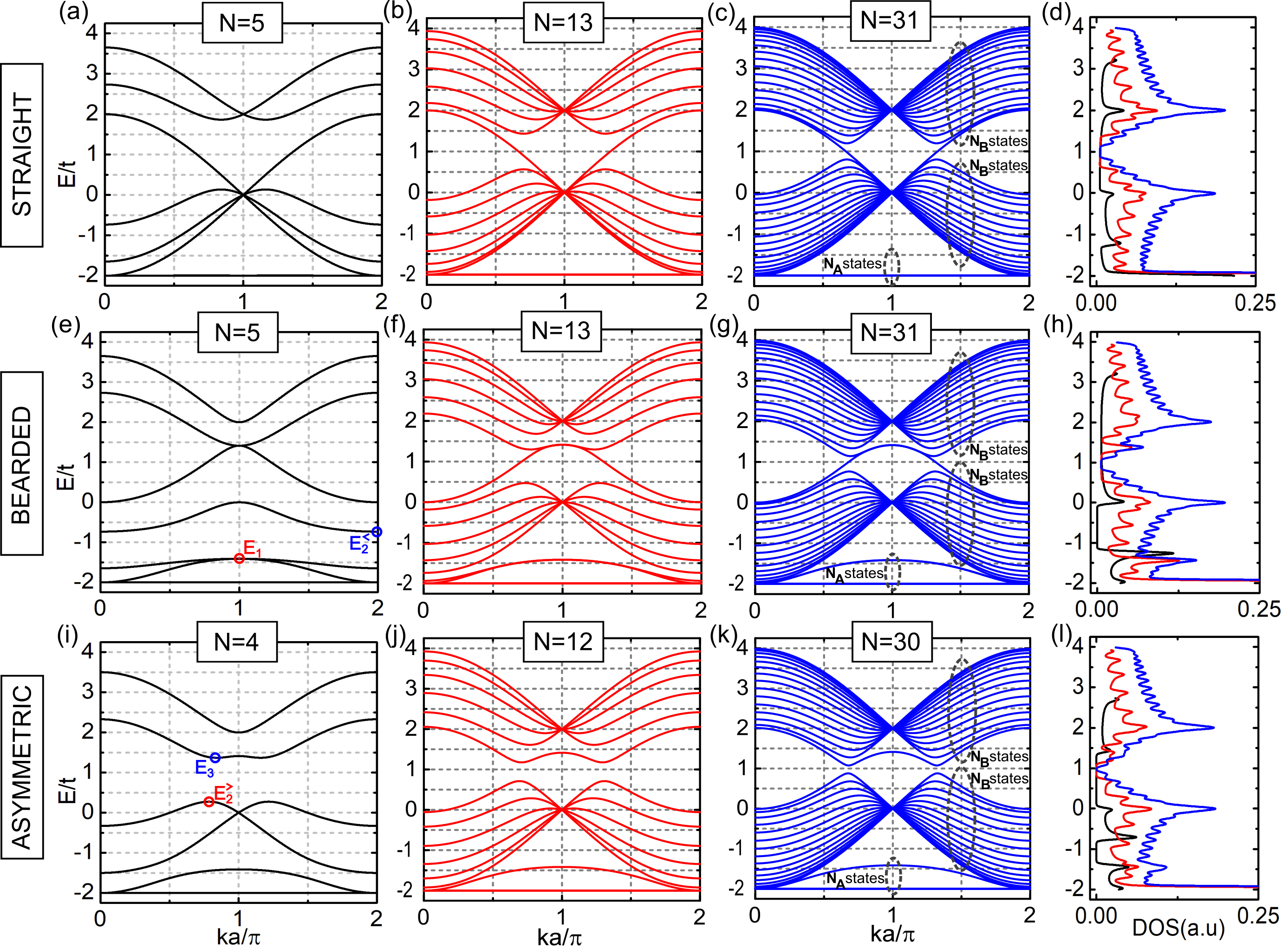}}
\caption{(Color online) The same as in Fig.~\ref{Fig2}, but now for kagome nanoribbon ($\theta=120^{\circ}$).}
	\label{Fig6}
\end{figure*}

From Figs.~\ref{Fig3} and \ref{Fig5}, it is clearly evident the general nodal-like behavior for the quasi-flat subbands for both Lieb and transition nanoribbons, regardless of the edge type. Within this context, an apparent difference in the wave function distributions for the most energetic state ($p=1$) of the quasi-flat subband is observed when comparing states (6) in Fig.~\ref{Fig3}(b) and state (4) in Fig.~\ref{Fig5}(b) for bearded edges, where in the former the wave function is equally distributed along whole nanoribbon in A sites, and in the latter it presents a well-defined peak in the central region of the nanoribbon. This apparent discrepancy also emerges from the choice of the momentum value for the plots, such that for the former, the $k$-point is in a region where the energy levels are, in fact, flat (see red asterisks (6) in the Fig.~\ref{Fig3}(b)), while in the latter, the electronic state presents a linear slope (see red asterisks (4) in the Fig.~\ref{Fig5}(b)), leading to a null and non-null group velocity, respectively.

Similarly to states (7) and (8) in Fig.~\ref{Fig3}(b) for bearded Lieb nanoribbons, bulk states (5) and (6) in Fig.~\ref{Fig5}(b) for bearded transition nanoribbons exhibit a clearer peak in the central region of the nanoribbon the larger is the momentum value far way of the ``anti-crossing'' region between the upper subband and quasi-flat subband. This can be noticed by comparing the decrease in the contribution of $|\psi_A|^2$ amplitudes of the total probability density between state (5) for $ka=\pi+0.1$ and state (6) for $ka=1.2\pi$ in Fig.~\ref{Fig5}(b) [see labeled states (5) and (6) in the inset of Fig.~\ref{Fig5}(b)], as well as between state (7) for $ka=\pi+0.1$ and state (8) for $ka=1.2\pi$ in Fig.~\ref{Fig3}(b) [see labeled states (7) and (8) in the inset of Fig.~\ref{Fig3}(b)]. An equivalent behavior of the $|\psi_A|^2$ decreasing contribution in the total probability density is also observed for the first energetic state of the conduction band for the asymmetric Lieb and transition nanoribbons, as respectively shown in panels (6) and (7) in Fig.~\ref{Fig3}(c) [see labeled states (6) and (7) in the inset of Fig.~\ref{Fig3}(c)] and in panel (6) in Fig.~\ref{Fig5}(c) [see labeled state (6) in the inset of Fig.~\ref{Fig5}(c)], presenting in both nanoribbon cases an asymmetric peak shifted to the straight edge side due to the chosen energetic mode with positive group velocity.

\begin{figure*}[t]
\centering
{\includegraphics[width=0.92\linewidth]{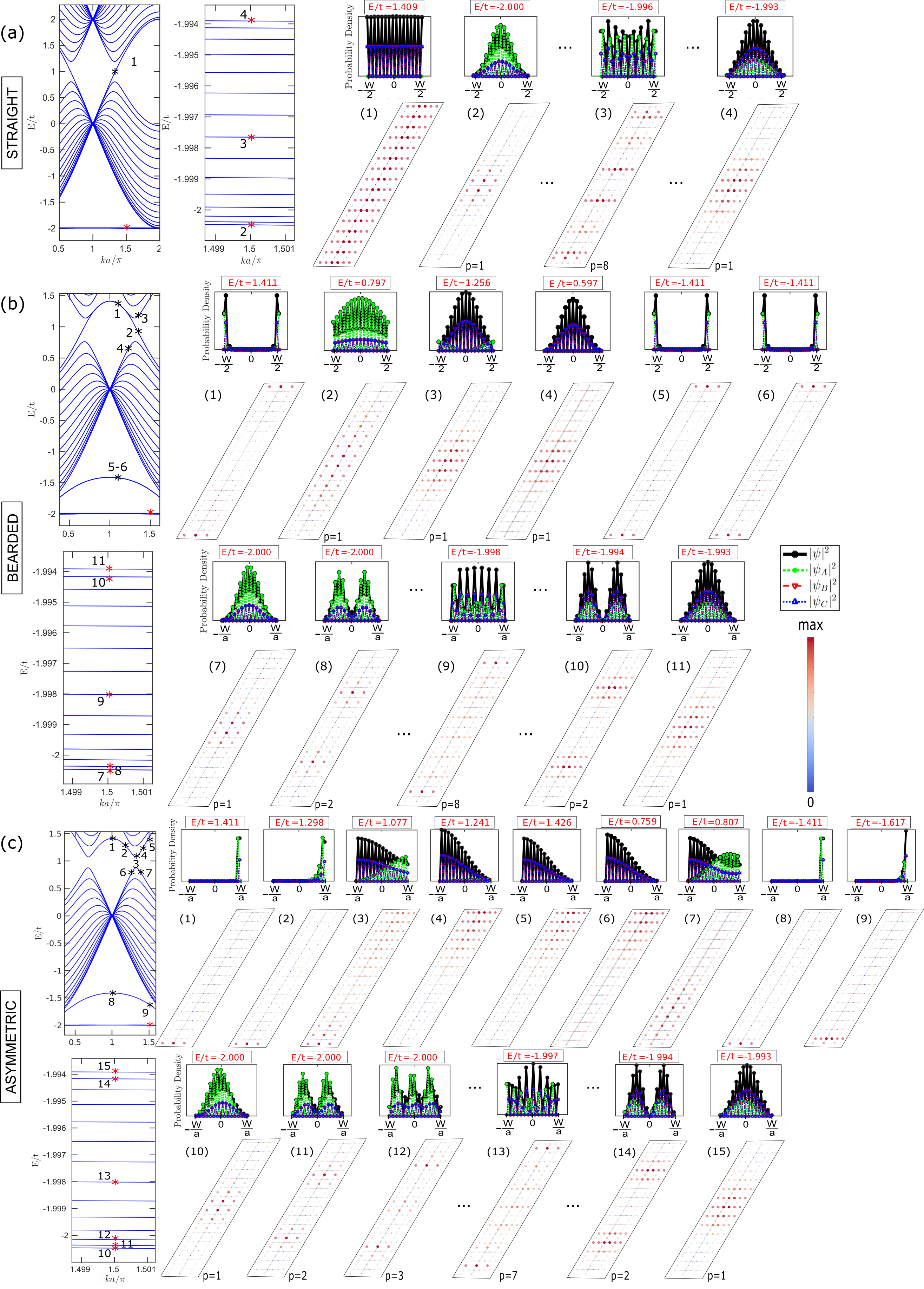}}
\caption{(Color online) The same as in Fig.~\ref{Fig3}, but now for a kagome nanoribbon ($\theta=120^\circ$). (a) Dirac-like state (1) and quasi-flat states (2)-(4) for the straight-edged case were calculated at $ka=4.228$ and $ka=3\pi/2$, respectively. (b) Edge states (1) and (5)-(6) for the bearded-edged case were calculated at $ka=\pi$, while for bulk-like states (2)-(3), (4), and (7)-(11) [see red asterisks in the bottom panel in (b)] were computed at $ka=4.312$, $ka=\pi+\pi/3$, and $ka=3\pi/2$ respectively. (c) The edge state (1) of the asymmetric-edged case was calculated at $ka=\pi$, and its evolution (2)-(5) was calculated for $ka=7\pi/6$, $ka=4\pi/3$, $ka=4.455$ and $ka=3\pi/2$, respectively. The bulk states (6)-(7) around the anti-crossing at $E/t\approx 1$ with opposite group velocities and consequently opposite edge localizations of the one-peak states were calculated for $ka=\pi+\pi/5$ and $ka=\pi+\pi/3$, respectively. The edge state of the bottom subband was shown at two different $k$-points (8) $ka=\pi$ and (9) $ka=3\pi/2$. Quasi-flat states in the bottom subband labeled by (10) to (15) [see red asterisks in the bottom panel in (c)] were calculated at $ka=3\pi/2$.}
\label{Fig7}
\end{figure*}

Finally, let us now discuss the results for kagome nanoribbons ($\theta=120^{\circ}$) with the three different edge terminations analyzed here: straight, bearded, and asymmetric. The band structures, the density of states, and the probability densities for different ribbon widths are respectively shown in Figs.~\ref{Fig6} and \ref{Fig7}. Since the nanoribbon energy spectra present discretized energy levels originating from the energetic bands of the bulk system, and considering that the band structure of the infinite sheet system of the kagome lattice is composed of a Dirac cone at $E=0$ and a flat band at $E=-2t$, in this way, one observes similarly in Fig.~\ref{Fig6} for kagome nanoribbons the formation of three energetic subbands with analog characteristics of the coexistence of a Dirac-like and quasi-flat states, as can be seen by the $N_A$ quasi-flat states formed around $E=-2t$ and the two subbands composed by $N_B$ bulk-like states. A similar state counting with one $N_A$ and two $N_B$ energetic subbands was also obtained for the cases of Lieb and transition nanoribbons as shown in Figs.~\ref{Fig2} and \ref{Fig4}, respectively.

Concerning the metallic or the semiconductor features depending on the edge termination type and ribbon width, one can notice from Fig.~\ref{Fig6} that kagome nanoribbons with straight edges are metallic [Figs.~\ref{Fig6}(a)-\ref{Fig6}(c)], similar to Lieb [Figs.~\ref{Fig2}(a)-\ref{Fig2}(c)] and transition [Figs.~\ref{Fig4}(a)-\ref{Fig4}(c)] nanoribbons with the same edge type, presenting such metallic nature, regardless the $\theta$ value, associated with the perfect edge termination with no dangling bonds likely in a infinite Lieb-kagome sheet. Kagome nanoribbons with bearded edges [Figs.~\ref{Fig6}(e)-\ref{Fig6}(g)] exhibit an energy gap between the higher-energetic level of the lower quasi-flat subband [$E_1$ - see red circle in Fig.~\ref{Fig6}(e) at $ka=\pi$] and the lower-energetic level of the dispersive subband [$E_2^<$ - see blue circle in Fig.~\ref{Fig6}(e) at $ka=2\pi$] only for $N \leq 7$, being $N=8$ the critical $N$ value that leads to the semiconductor-to-metallic transition for bearded kagome nanoribbons, as shall be discussed further in Sec.~\ref{Sec.laws_gap} [see blue circle for $\Delta_{12}^B$ in Fig.~\ref{Fig9}(c)]. For $N \geq 9$, the minimum of the middle subband ($E_2^<$) lies at a lower energy value than the maximum of the lower subband ($E_1$) at a different momentum, \textit{i.e.} $E_1 > E_2^<$, such that there is no full energy gap in the energy spectrum of the bearded-edged kagome nanoribbon, although there is no crossing between the energy states of middle and upper subbands being separated at each point in $k$ space, as can be verified for $N=13$ in Fig.~\ref{Fig6}(f). In a similar energetic situation of no full gap and non-touching bands, but analyzing the band structure of the bulk system, Ref.~[\onlinecite{belgeling2012}] referred to this band configuration, saying that the spectrum exhibits a negative indirect gap. In this example, a \(1/3\)-filled system will always have both the lower subband and the middle band partially filled. Similarly, a \(2/3\)-filled system will always have both the upper band and the middle band partially filled, indicating a semimetallic regime. For \(2/3\)-filling, regardless of \(N\), there is no gap, and the regime is metallic, differing from Lieb [Figs.~\ref{Fig2}(e-g)] and transition [Figs.~\ref{Fig4}(e-g)] nanoribbons with bearded edges, which are semiconductors for a \(2/3\)-filling. Kagome nanoribbons with asymmetric edges [Figs.~\ref{Fig6}(i)-\ref{Fig6}(k)] are semiconductors, similar to the equivalent cases in Lieb [Figs.~\ref{Fig2}(i)-\ref{Fig2}(k)] and transition [Figs.~\ref{Fig4}(i)-\ref{Fig4}(k)] nanoribbons with the same type of edges. However, unlike asymmetric-edged Lieb and transition nanoribbons, which exhibit two energy gaps between the dispersive bottom and upper subbands and the quasi-flat middle subband, asymmetric-edged kagome nanoribbons present only one energy gap between the higher-energetic state in the dispersive subband [$E_2^>$ - see red circle in Fig.~\ref{Fig6}(i)] and the lower-energetic state in the upper dispersive subband [$E_3$ - see blue circle in Fig.~\ref{Fig6}(i)].

The corresponding density of states for the three edge terminations of the investigated kagome nanoribbons are depicted in Fig.~\ref{Fig6}(d) for straight, Fig.~\ref{Fig6}(h) for bearded, and Fig.~\ref{Fig6}(l) for asymmetric boundaries, in which the different colors are associated with the different ribbon widths in the first, second, and third columns of panels in Fig.~\ref{Fig6}. Due to the degenerate flat states at $E=-2t$ in the energy spectra of the three investigated types of kagome nanoribbons, one observes a pronounced peak in the density of states at this energy value. It is important to mention that the position of the flat band is located at the bottom or top of the energy spectrum in the kagome lattice when one assumes the hopping to be positive ($t > 0$) or negative ($t < 0$) in the tight-binding model,\cite{tony2019, lima2022} and consequently, the position of the corresponding van Hove singularity in the density of states will be localized at $E=-2t$ or $E=2t$ with a shifted spectrum for $t > 0$ or $t < 0$. From Figs.~\ref{Fig6}(d), \ref{Fig6}(h), and \ref{Fig6}(l), one notices two pronounced peaks in the density of states at the energies $E=+2t$ and $E=0$, arising from the highly degenerate states at $ka=\pi$ for the three different edges. Additionally, bearded and asymmetric kagome nanoribbons exhibit an additional peak in the density of states around $E\approx -1.5t$, which originates from double-degenerate (single) states in the bearded (asymmetric) case. These states compose the $N_A$ quasi-flat subband of the kagome nanoribbon, being in close proximity to the other flat states at $ka=0$ and $ka=2\pi$, but deviate from them as $ka$ approaches $\pi$.

Beyond the energetic aspect of the dispersion relations discussed here for each of the three types of edge terminations for the kagome nanoribbons, in order to distinguish between bulk, edge, and quasi-flat states, we show in Fig.~\ref{Fig7} the probability densities for [Fig.~\ref{Fig7}(a)] straight, [Fig.~\ref{Fig7}(b)] bearded, and [Fig.~\ref{Fig7}(c)] asymmetric kagome nanoribbons, presenting a cross-section along the nanoribbon and in a perspective view in the top and bottom panels at each row of plots, respectively. Similar to the cases of Lieb nanoribbons, shown in panel (6) of Fig.~\ref{Fig3}(a), and transition nanoribbons, shown in panel (6) of Fig.~\ref{Fig5}(a), the Dirac-like state labeled by (1) in Fig.~\ref{Fig7}(a) for straight-edged kagome nanoribbon corresponds to a bulk state exhibiting an equally distributed probability density along the whole nanoribbon with no-null contributions coming from $|\psi_B|^2$ and $|\psi_C|^2$. Moreover, the probability densities of the $N_A$ states from the quasi-flat subbands exhibit a typical nodal behavior, such that the peak counting is given by $p\in [1,2,\cdots,(N_A-1)/2,N_A/2,(N_A-1)/2,\cdots,2,1]$ for the straight-edged kagome nanoribbons. Panels (2) to (4) in Fig.~\ref{Fig7}(a) exemplify the modes with $p=1$ and $p=8$ peaks. Due to the symmetric edge configuration, the quasi-flat states for straight-edged kagome nanoribbons exhibit peaks' distributions in a symmetrical manner, as can be seen for the lowest energy quasi-flat state in panel (2) of Fig.~\ref{Fig7}(a) with a symmetrical peak with respect to the central axis of the nanoribbon. Similar to the spatial distribution of the quasi-flat states in straight-edged Lieb nanoribbons [see panels (1) to (5) in Fig.~\ref{Fig3}(a)], for straight-edged kagome case [see panels (2) to (4) in Fig.~\ref{Fig7}(a)], one also notices an inversion of the sublattices contributions to the total wave functions of the quasi-flat states when one analyzes states with $p \leq N_A/2$ and $p> N_A/2$. Here, for the kagome case, the states display higher intensities at type A sublattices and lower and equal intensities at B and C sublattices for the first half of quasi-flat states, \textit{i.e.} $p\in [1,2,\cdots,(N_A-1)/2,N_A/2]$, and an inversion on the probability density contribution for the second half of quasi-flat states, \textit{i.e.} $p\in [(N_A-1)/2,\cdots,2,1]$, with higher and equal intensities at B and C sublattices and lower intensities at type A sublattices. Unlike the straight-edged Lieb case, here, for the straight-edged kagome nanoribbons, all sublattices A, B, and C have non-null probability densities, whereas for the Lieb case, $|\psi_B|^2=0$ for any $p$ mode in the quasi-flat subband.

In bearded kagome nanoribbons [Fig.~\ref{Fig7}(b)], four edge states are observed: two of them around $E/t \approx 1$, arising in the upper-middle band-touching, and the other two edge states around $E/t \approx -1.5$, coming from the quasi-flat subband. Panel (1) in Fig.~\ref{Fig7}(b) for the state highlighted by (1) in the dispersion relation depicts the probability density of the set of the two edge states with energy around $E/t \approx 1$, showing the typical surface behavior with an exponentially decaying mode with a high concentration of the probability density localized at the edges. Far from the degeneracy region ($ka/\pi\lesssim 0.7$ and $ka/\pi \gtrsim 1.3$) between the two edge states around $E/t \approx 1$, a mixed edge-bulk-like state contribution between the edge state and the states from the nearest bulk dispersive subbands are verified, owing to the anti-crossing and the degeneracy breaking of the edge states. This can be seen in the states (2) and (3) in Fig.~\ref{Fig7}(b), which exhibit probability densities with a mix of one peak, such as a $p=1$ bulk dispersive state, and an edge contribution (see green symbols denoting the $|\psi_A|^2$ amplitude with edge contribution). Note that, a bit further $k$-value away from the anti-crossing, the state (4) of the middle subband has the same pattern as the state (3) of the upper subband in Fig.~\ref{Fig7}(b), agreeing with the fact that they should have a similar spatial wave function distribution in case that the anti-crossing would be absent, as observed in other confinement nanostructures as reported for graphene quantum dots in Ref.~[\onlinecite{lavor2020magnetic}], which analyzed the crossing and anti-crossings of the confined states. Panels (5) and (6) in Fig.~\ref{Fig7}(b) show the probability density for the other two edge states with energy values around $E/t \approx -1.5$ in the quasi-flat subband. These states are double-degenerate at every $k$-value and have their probability densities located mainly on the edge sites with dangling bonds, being the bearded edges here formed by A sublattices as sketched in Fig.~\ref{Fig1}(c) and as also verified by the higher intensities in A sites denoted by green symbols in the cross-section panels (1), (5) and (6) in Fig.~\ref{Fig7}(b). Similarly to the straight-edged kagome nanoribbons [panels (2) to (4) in Fig.~\ref{Fig7}(a)], the quasi-flat states of bearded-edged kagome nanoribbons also display a sublattice contribution inversion in the total probability density roughly between the first and second half part of quasi-flat states. It can be checked in panels (7) to (11) in Fig.~\ref{Fig7}(b) [marked as red asterisks] where the nodal states have higher intensities at A (B and C) sublattices for approximately the first (second) half of the quasi-flat states. However, unlike the bulk-like nodal quasi-flat states of the straight-edged kagome nanoribbons and Lieb and transition nanoribbons with any type of edge, the $N_A-2$ flat states in Fig.~\ref{Fig7}(b) do not exhibit a continuous increase in the number of peaks, as can be noticed by the seventh flat state marked as the point (9) in Fig.~\ref{Fig7}(b) that displays eight peaks ($p=8$) instead of the expected seven peaks. Therefore, the bottom subband for bearded-edged kagome nanoribbons is composed of two edge states and $N_A-2$ flat states with nodal bulk-like behavior, totalizing $N_A=(N+1)/2$ states. To our knowledge, such edge states [panels (1), (5), and (6) in Fig.~\ref{Fig7}(b)] had not been reported in the literature for the bearded-edged kagome case in the absence of ISO coupling.

Kagome nanoribbons with asymmetric edges [Fig.~\ref{Fig7}(c)] exhibit two edge states. Unlike the edge states for the bearded-edged kagome case, the asymmetric kagome nanoribbons' edge states had already been reported in the literature, as discussed in Ref.~[\onlinecite{wang2008edge}]. Due to its crystallographic configuration with the non-symmetric boundaries, formed by bearded and straight terminations, edge states in asymmetric-edged kagome nanoribbons display a preferential localization in the bearded side of the nanoribbon owing to the presence of the dangling bonds, as shown in panels (1) and (2) in Fig.~\ref{Fig7}(c) for the first state of the upper subband with two different $k$ values, and in panels (8) and (9) in Fig.~\ref{Fig7}(c) for the highest energetic state in the bottom subband with two different $k$ values. Note that in the case of the asymmetric edge, there is only one edge state originating in the bottom subband [states (8) and (9) in Fig.~\ref{Fig7}(c)], while for the bearded-edged kagome case, this is a double-degenerate state [states (5) and (6) in Fig.~\ref{Fig7}(b)]. This is linked to the number $N_A$ of modes in the quasi-flat subband, that in the bearded case ($N_A=(N+1)/2$) has one more sublattice A than the asymmetric case ($N_A=N/2$). To emphasize the anti-crossing effect of mixing the mode as a $p=1$-bulk and edge-like state, we show in panels (3) to (5) and in panels (6) and (7) the probability densities for the lowest state in the upper subband and the highest state in the middle subband, respectively, with different $k$ values. Note that the state (3) closest to the anti-crossing point exhibits this mixing in its probability density formed by an edge state with higher intensities at A sites (green symbols - $|\psi_A|^2$) and an asymmetric peak pushed to the straight termination with higher intensities at B and C sublattices (red and blue symbols - $|\psi_B|^2$ and $|\psi_C|^2$). The probability density evolution of the edge state with the lowest energy in the upper subband for different $k$-values in panels (1) to (5) in Fig.~\ref{Fig7}(c) shows a localization transition initially from the bearded side of the nanoribbon [panels (1) and (2) in Fig.~\ref{Fig7}(c)] to the straight side of the nanoribbon [panels (4) and (5) in Fig.~\ref{Fig7}(c)], that can be verified by the group velocity sign inversion when one analyzes the energetic edge state derivative. As expected due to the anti-crossing, state (6) in the middle subband has an equivalent probability density pattern as in states (4) and (5) in the upper subband in Fig.~\ref{Fig7}(c) with positive group velocities, as well as the correspondence between state (3) and state (7) with negative group velocities. Similar to the $N_A$ quasi-flat states of asymmetric-edged Lieb [see states (1) to (5) in Fig.~\ref{Fig3}(c)] and transition [see states (1) to (5) in Fig.~\ref{Fig5}(c)] nanoribbons, the flat states in asymmetric-edged kagome nanoribbons display a nodal behavior with roughly asymmetric wave function distribution with respect to the central of the nanoribbon, as shown in panels (10) to (15) in Fig.~\ref{Fig7}(c), but now composed by $N_A-1$ states.

\begin{figure*}[t]
\centering
{\includegraphics[width=\linewidth]{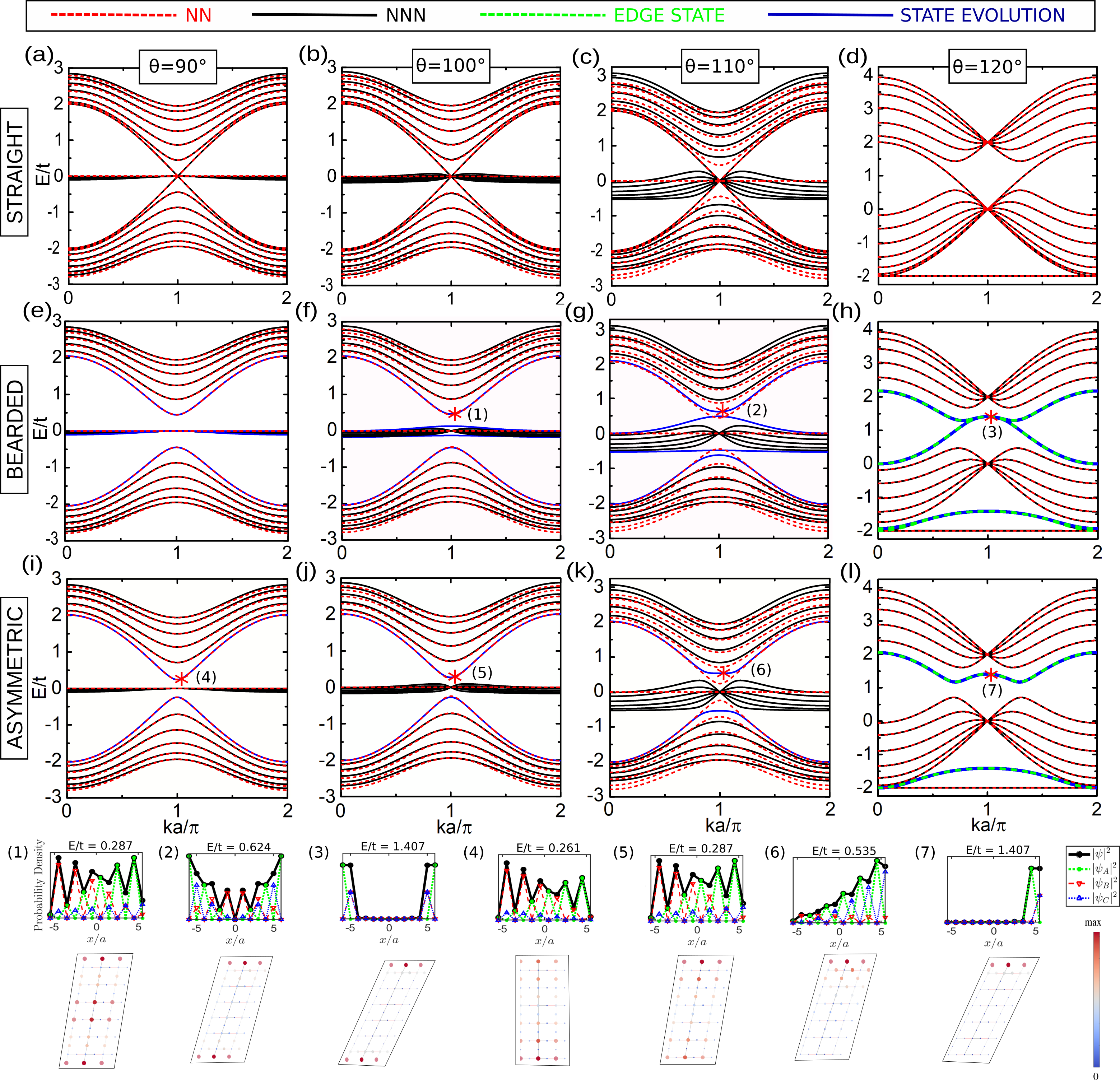}}
\caption{(Online color) Energy spectra obtained within the tight-binding approximation taking into account the nearest-neighbor (NN) connections (red dashed curves) and next-nearest-neighbor (NNN) connections (black solid curves) for different $\theta$ parameters and different types of terminations: Lieb ($\theta=90^{\circ}$ - first column), transition ($\theta=100^{\circ}$ - second column and $\theta=110^{\circ}$ - third column), and kagome ($\theta=120^{\circ}$ - fourth column) nanoribbons with straight edges ($N=13$ - first row), bearded edges ($N=13$ - second row), and asymmetric edges ($N=12$ - third row). The evolution of the energetic levels in the formation of the edge states when achieves $\theta=120^\circ$ is highlighted by blue curves. Green dashed curves in panels (h) and (l) correspond to the edge states. The bottom panels depict the probability density distributions within the NNN approach for the states labeled by (1) to (7) in panels (f) to (l) to illustrate the edge state formation in the interconvertibility process between Lieb and kagome nanoribbons. From the selected states, only states (3) and (7) for kagome nanoribbons correspond to edge states.}
\label{Fig8}
\end{figure*}

\subsection{Effects of next-nearest-neighbor sites and state evolution into edge states}\label{Sec.NNN_sites}

The results presented in the previous section were obtained using a tight-binding model that incorporates all hopping parameters depicted in Fig.~\ref{Fig1}(a) and governed by Eq.~\eqref{hopping}. In this section, we examine the effects of considering non-zero hoppings only between NN sites, specifically $t_{BA}$ and $t_{BC}$ for Lieb and transition nanoribbons, and $t_{BA}$, $t_{BC}$, and $t_{AC}$ for kagome nanoribbons. We refer to this model as the NN-sites approximation for comparison purposes, while the model utilized in the previous section will be referred to as the NNN-sites approximation. The effective inclusion of hopping parameters beyond nearest neighbors, such as NNN and third neighbors, is only feasible when $n<8$ in Eq.~\eqref{hopping}, as demonstrated in Appendix \ref{sec.appendix_A}.

\begin{figure*}[!t]
\centering
{\hspace{-0.5cm}\includegraphics[width=1.013\linewidth]{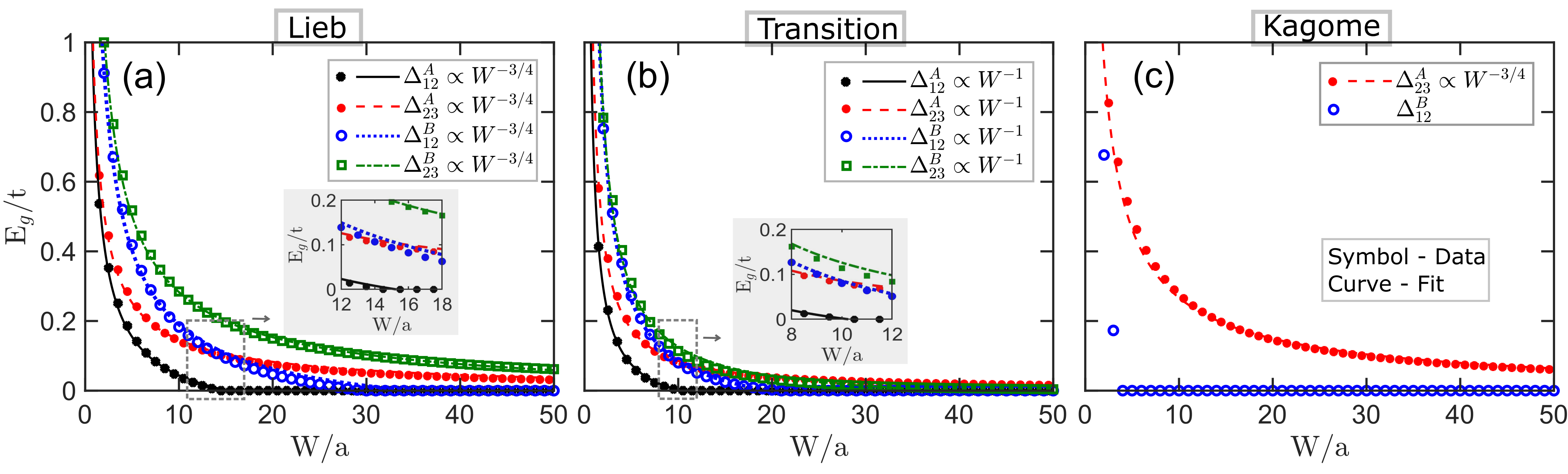}}
\caption{(Color online) Band gap energies as a function of the nanoribbon width, $W=(N-1)a_0$, for: (a) Lieb, (b) transition and (c) kagome nanoribbons. $\Delta_{12}$ ($\Delta_{23}$) denotes the energetic difference between the highest energy of the lower (middle) subband and the lowest energy of the middle (upper) subband. These energies are indicated in Figs.~\ref{Fig2}(e) and \ref{Fig2}(i) for Lieb nanoribbons and in Figs.~\ref{Fig4}(e) and \ref{Fig4}(i) for transition nanoribbons, taking asymmetric (A-superscript) and bearded (B-superscript) edges, respectively. Symbols (lines) correspond to the data points (fit curves). The insets in panels (a) and (b) show enlargements for the regions denoted by the rectangular gray box to highlight the crossing between the energetic differences $\Delta_{23}^A$ and $\Delta_{12}^B$.}
\label{Fig9}
\end{figure*}

Figure~\ref{Fig8} illustrates the dispersion relation evolution of nanoribbons with the arrangement of Lieb lattice ($\theta=90^{\circ}$ - $1^{\text{st}}$ column), transition lattices with $\theta=100^{\circ}$ ($2^{\text{nd}}$ column) and $\theta=110^{\circ}$ ($3^{\text{rd}}$ column) and kagome lattice ($\theta=100^{\circ}$ - $4^{\text{th}}$ column). Different types of edges are analyzed: straight edges [Figs.~\ref{Fig8}(a)-\ref{Fig8}(d)], bearded edges [Figs.~\ref{Fig8}(e)-\ref{Fig8}(h)], and asymmetric edges [Figs.~\ref{Fig8}(i)-\ref{Fig8}(l)]. The results obtained using the NN-sites and NNN-sites approximations are represented by red dashed and black solid curves, respectively. Note by the energy spectra in red dashed curves, when considering only NN-sites, in Figs.~\ref{Fig8}(a)-\ref{Fig8}(c), \ref{Fig8}(e)-\ref{Fig8}(g), and \ref{Fig8}(i)-\ref{Fig8}(k) for straight, bearded, and asymmetric edge cases, respectively, that there are no pronounced changes in the energy spectra in the interconvertibility process during the lattice morphologic transition from Lieb to transition lattice regardless the edge termination type. For instance, for all three edge types, the middle flat subband remains $N_A$th-fold degenerate and fully flat in the whole $k$-range. On the other hand, by considering connections between NNN-sites (solid black curves), one breaks the degeneracy of the states that form the middle subband, increasing the dispersivity of the middle subband, leading, in turn, to the flat subband becoming quasi-flat subband and consequently to an increase in its bandwidth the high the $\theta$-parameter in the transition stages. This non-change of the middle subband during the interconvertibility process is due to the chosen value for the $n$-parameter as $n=8$ in Eq.~\eqref{hopping}, which reduces long-range interactions. It is important to mention that for $n<8$, as discussed in Appendix \ref{sec.appendix_A}, these subbands become increasingly dispersive. In other words, the identical energy spectrum for transition and Lieb nanoribbons in the NN-sites approximation is because the NN-sites model does not consider the effects of the approximation's diagonal deformation of the closest A and C sites since $t_{AC}$ is assumed to be zero. Only when $\theta=120^{\circ}$ the spectrum of the NN-sites approximation abruptly change to the energy spectrum of the kagome nanoribbon, as the hopping parameter $t_{AC}$ becomes NN-sites parameter in kagome nanoribbons, instead of an NNN-sites parameter as in the transition nanoribbons.

To guide the eyes on the Lieb-to-kagome evolution of the states that shall correspond to the edge states in the kagome nanoribbon case, we highlighted the state evolution in blue in Figs.~\ref{Fig8}(e)-\ref{Fig8}(g) for bearded nanoribbons and in Figs.~\ref{Fig8}(i)-\ref{Fig8}(k) for asymmetric nanoribbons. As discussed in the previous section, bearded-edged kagome nanoribbons have four edge states, as highlighted in green dashed curves in Fig.~\ref{Fig8}(h). According to Figs.~\ref{Fig8}(e)-\ref{Fig8}(h), the two edge states with energy value between $0\lessapprox E/t \lessapprox 2$ for kagome case in Fig.~\ref{Fig8}(h) emerge from the highest energy state of the middle subband and the lowest energy state of the upper subband, whereas the other two edge states that are double-degenerate and that one forms at the bottom subband in the kagome nanoribbon energy spectrum emerge from the highest energy state of the bottom subband and the lowest energy state of the middle subband. For asymmetric nanoribbons, Figs.~\ref{Fig8}(i)-\ref{Fig8}(l) elucidate that the two edge states in the kagome nanoribbon energy spectrum [Figs.~\ref{Fig8}(l)] emerge from the highest and lowest energy states of the bottom and upper dispersive subbands, respectively. One of the edge states for asymmetric kagome nanoribbons is energetically localized at the middle gap, and the other is the highest energetic state of the bottom subband. Interestingly, bearded nanoribbons have one more state in the quasi-flat subband $(N_A=(N+1)/2)$ than the other two straight $(N_A=(N-1)/2)$ and asymmetric $(N_A=N/2)$ edge cases, and it is the only case in which the quasi-flat subband presents a two-state degeneracy breaking at $ka/\pi=1$. These facts are associated with the reason why, for bearded kagome nanoribbons, one has the formation of two pairs of edge states, one of them double-degenerates, whereas, in the asymmetric kagome nanoribbons, one has solely two edge states and they are not degenerate.

As one example of an edge state evolution, we also present at the bottom of Fig.~\ref{Fig8} the probability densities for the states labeled by (1) to (3) for the bearded edge case and by (4) to (7) for the asymmetric edge case. From states (1) and (3), one notices the transition of the bulk-like state [panels (1) and (2)] into an edge state [panel (3)] for bearded nanoribbons. A similar bulk-to-edge tendency of the states (4) to (7) is also observed for asymmetric nanoribbons. Therefore, we demonstrated here that asymmetric and bearded kagome nanoribbons present edge states not only in the NNN-sites approximation but also in the NN-sites approximation and that by analyzing the consequences on the energy spectrum during the Lieb-kagome morphology within the NNN-sites approach, we can identify the states evolution into the edge states.

\subsection{Scaling laws of band gap energies}\label{Sec.laws_gap} 

The size dependence of the band gap is a crucial physical feature for practical optical applications of quantum confinement systems for designing new nanoelectronic devices. In view of the relevance of this context, we also investigate here the scaling laws governing the band gap energies of Lieb, transition, and kagome nanoribbons using the tight-binding model described in Sec.~\ref{sec.II.B}. A similar investigation has been reported in the literature for nanoribbons made out of other 2D materials, for instance, graphene nanoribbons \cite{son2006energy, brey2006electronic, wakabayashi2010electronic} and phosphorene nanoribbons \cite{wu2015electronic, de2016boundary, sisakht2015, tran2014scaling}. For that, we systematically searched in the first Brillouin zone, by varying $k_x$ and $k_y$ from 0 to $2\pi$, and calculated $\Delta_{12} = |E_2^< - E_1|$ and $\Delta_{23} = |E_3 - E_2^>|$, which are the smallest energy difference between the highest energetic mode of the lower subband ($E_1$) and the lowest energetic mode of the middle subband ($E_2^<$), and the smallest energy difference between the highest energetic mode of the middle subband ($E_2^>$) and the lowest energetic mode of the upper subband ($E_3$), respectively. Such energy levels $E_1$, $E_2^<$, $E_2^>$ and $E_3$ are depicted as an example case in Figs.~\ref{Fig2}(e) and \ref{Fig2}(i) for Lieb, \ref{Fig4}(e) and \ref{Fig4}(i) for transition, and \ref{Fig6}(e) and \ref{Fig6}(i) for kagome nanoribbons for the smallest possible nanoribbon width with bearded and asymmetric edges, respectively. Due to the metallic nature of straight nanoribbons, \textit{i.e.}, with an energy band gap absent, our analysis becomes restricted only to bearded and asymmetric-edged nanoribbons, where the energetic differences are denoted with a B and an S superscript such as $\Delta^B_{ij}$ and $\Delta^S_{ij}$, respectively. In Fig.~\ref{Fig9}, we show the behavior of the energy band gap as a function of the ribbon width for [Fig.~\ref{Fig9}(a)] Lieb, [Fig.~\ref{Fig9}(b)] transition, and [Fig.~\ref{Fig9}(c)] kagome nanoribbons. Symbols correspond to the obtained tight-binding data values, and the curves are power-law fitting curves.

In general, the dependence of the energy gaps on the width ($W$) of asymmetric ($\Delta_{ij}^A$) and bearded-edged ($\Delta_{ij}^B$) Lieb nanoribbons follows a power law $\propto W^{-3/4}$ [Fig.~\ref{Fig9}(a)]. It is worth noting that $\Delta^{A,B}_{23}>\Delta^{A,B}_{12}$ for all values of $W$ in both edge configurations, and that $\Delta_{12}^A$ and $\Delta_{12}^B$ decay faster to zero than $\Delta_{23}^A$ and $\Delta_{23}^B$, becoming zero for $W^A\geq14.5a$ and $W^B\geq30a$, respectively. This difference between the energy gaps $\Delta^{A,B}_{12}$ and $\Delta^{A,B}_{23}$ is due to the positive-negative asymmetry of the Lieb energy spectra caused by the bandwidth of the quasi-flat subband which forms mainly in the region below $E/t=0$ at the corners of the Brillouin zone at $ka/\pi=0$ and $ka/\pi=2$ [see Fig.~\ref{Fig2}]. Furthermore, bearded Lieb nanoribbons exhibit $\Delta_{12}^B>\Delta_{23}^A$ for $W\leq13a$, and $\Delta_{12}^B<\Delta_{23}^B$ for $W>13a$ [see inset in Fig.~\ref{Fig9}(a)].

Unlike the Lieb case, the corresponding gaps in the energy spectrum of transition nanoribbons with $\theta=105^\circ$ [Fig.~\ref{Fig9}(b)] exhibit a dependence on the width proportional to $W^{-1}$ for both asymmetric and bearded edges. Since the bandwidth of the transition nanoribbons is greater than the Lieb nanoribbons, the energy gaps for the transition nanoribbons in Fig.~\ref{Fig9}(b) are smaller than the Lieb cases in Fig.~\ref{Fig9}(a), as well as the energetic difference between $\Delta_{12}^{A,B}$ and $\Delta_{23}^{A,B}$ for transition nanoribbons are smaller than the Lieb cases. Furthermore, $\Delta_{12}^A$ and $\Delta_{12}^B$ in Fig.~\ref{Fig9}(b) become zero for $W\geq10.5a$ and $W\geq20a$, respectively. The inset in Fig.~\ref{Fig9}(b) shows an enlargement around $W/a\approx 10$, presenting a crossing between $\Delta_{12}^B$ and $\Delta_{23}^A$, such that $\Delta_{12}^B>\Delta_{23}^A$ for $W\leq9a$, and $\Delta_{12}^B<\Delta_{23}^A$ for $W>9a$.

From Figs.~\ref{Fig6}(e)-\ref{Fig6}(g) and following the evolution of the energy spectra for kagome nanoribbons with increasing width, one observes that the bearded-edged kagome nanoribbons exhibit a non-zero gap ($\Delta_{12}^B$) between bottom and middle subbands only for the two smallest nanoribbon sizes with values of $\Delta_{12}^B=0.677t$ for $N=5$ and $\Delta_{12}^B=0.173t$ for $N=7$, corresponding the two non-zero data denoted by blue opened circular symbols in Fig.~\ref{Fig9}(c). On the other hand, asymmetric-edged kagome nanoribbons [see Figs.~\ref{Fig6}(i)-\ref{Fig6}(k)] exhibit an energy gap ($\Delta_{23}^A$) between the middle and upper subbands that obeys a width-dependent relationship proportional to $W^{-3/4}$ [see red dashed curve in Fig.~\ref{Fig9}(c)], similar to the asymmetric- and bearded-edged Lieb cases in Fig.~\ref{Fig9}(a).

In analogy with graphene nanoribbons with armchair edges \cite{wakabayashi2010electronic, brey2006electronic}, it was to be expected at first glance that the Lieb and kagome nanoribbons with semiconductor character (asymmetric and bearded edges) should exhibit the energy gap being inversely proportional to the ribbon width, \textit{i.e.} $\propto W^{-1}$. Likely graphene, the band structures of both Lieb and kagome lattices are composed of a Dirac cone, and in addition here, by the coexistence of a flat band, with the Dirac dispersive bands being given within the long-wavelength regime by $E_{1,3}=\mp \hbar v_F k$ ($v_F$ is the Fermi velocity) for the bottom and upper bands for Lieb lattice and $E_{2,3}=t \mp \hbar v_F k$ for the middle and upper bands for kagome lattice. Thus, by assuming $k\propto n\pi/W$ ($n$ is an integer) for a nanoribbon with the length size $W$, one obtains $\Delta E \propto W^{-1}$ for the energy difference between the two dispersive subbands for semiconductor nanoribbons. However, this expected result obtained only by analyzing the bulk bands of an infinite system and the discretized $k$-values due to the finite-size effect does not give us, in fact, the entire physics of the bulk subband of Lieb-kagome nanoribbons, as can be seen by the width dependence of $W^{-3/4}$ for Lieb and Kagome nanoribbons and $W^{-1}$ for transition nanoribbons.

Our scaling law analysis on the dependence of the energy gaps with respect to the ribbon width reveals a different tendency for transition nanoribbons in the interconvertibility processes between Lieb and kagome nanoribbons, with a band gap decaying faster in the transition case than both Lieb and kagome nanoribbons, and that the bearded-edged Lieb and transition nanoribbons exhibit the larger band gaps, with the former case exhibiting the largest values. Therefore, our findings suggest that Lieb-kagome nanoribbons' electronic properties and gap engineering can be tailored by controlling the $\theta$-parameter and the edge type.

\section{Conclusion}\label{sec.IV}

In this work, we systematically investigated the electronic properties of nanoribbons composed of Lieb, transition, and kagome lattices using a tight-binding Hamiltonian that incorporates the interconvertibility between these lattices. Our analysis included the energy spectrum, density of states, and probability densities for nanoribbons with different edge terminations (straight, bearded, and asymmetric) and widths. We also explored the impact of NN and NNN interactions on the energy spectrum and on the evolution of the energy levels that make to originate the edge states in asymmetric and bearded-edged kagome nanoribbons. 

For all investigated nanoribbons, we verified that the number of quasi-flat states is equal to the number of lines that contain A sites $(N_A)$, while the number of states in the dispersive subbands corresponds to the number of B sites $(N_B)$. By exploring the probability densities of the states, we showed that both quasi-flat and dispersive states exhibit a bulk-like nodal behavior and that the Dirac-like state in the energy spectrum of nanoribbons with straight edges is not an edge state as one might expect but rather a bulk-like state. As a consequence of its crystallographic termination, we showed that bearded-edged nanoribbons exhibit one more state in the quasi-flat subband than straight and asymmetric nanoribbons. We demonstrated that only the energy spectra of kagome nanoribbons exhibit edge states: four edge states, two of them double-degenerate, for the bearded-edged case and two edge states for asymmetric terminations. This is in full agreement with the previously reported results in the literature \cite{zhang2016dispersion, chen2016finite} that showed in the absence of ISO coupling that Lieb nanoribbons do not have surface states. We also showed here that as well transition nanoribbons do not present surface states. Our results also demonstrated that straight nanoribbons exhibit metallic behavior, whereas Lieb and transition nanoribbons with bearded or asymmetric edges exhibit a semiconductor feature with two energy gaps in their spectra. For the kagome case, only the two smallest bearded-edged nanoribbons with $N=5$ and $N=7$ have an energy gap between the bottom and middle subbands. In contrast, asymmetric-edged kagome nanoribbons are semiconductors with a non-null energy gap for a wide range of ribbon width. 

Within the NNN-sites approach, we showed that transition nanoribbons have quasi-flat subbands that are more dispersive than those in Lieb and kagome nanoribbons, whereas, within the NN-sites approach, the flat subband remains fully flat in the whole Brillouin zone regardless of the edge type and the number of lines for Lieb and transition nanoribbon cases, and consequently leading to energy spectra with a positive-negative symmetry, showing that the dispersive energetic character of these bands results from the NNN interactions. Regarding the band gap energy dependence on the ribbon width, we found that the energy gaps of asymmetric and bearded-edged Lieb nanoribbons and asymmetric-edged kagome nanoribbons follow a power law dependence on the width of $\propto W^{-3/4}$. For transition nanoribbons assuming $\theta=105^{\circ}$, it exhibits a width-dependent behavior $\propto W^{-1}$ for both asymmetric and bearded edges.

The present systematic investigation offers insights into the electronic structure of Lieb-kagome nanoribbons, which could potentially be applied in the design of future nanoelectronic devices and in the exploration of nanostructures based on flat band materials. Moreover, the general theoretical framework present here allows for exploring the evolution stages between Lieb and Kagome nanoribbons and the consequences of their physical properties by means of just one control parameter.

\appendix

\section{The role of the $n$-parameter in the hopping energies}\label{sec.appendix_A}

\begin{figure}[t!]
\centering
{\includegraphics[width=\linewidth]{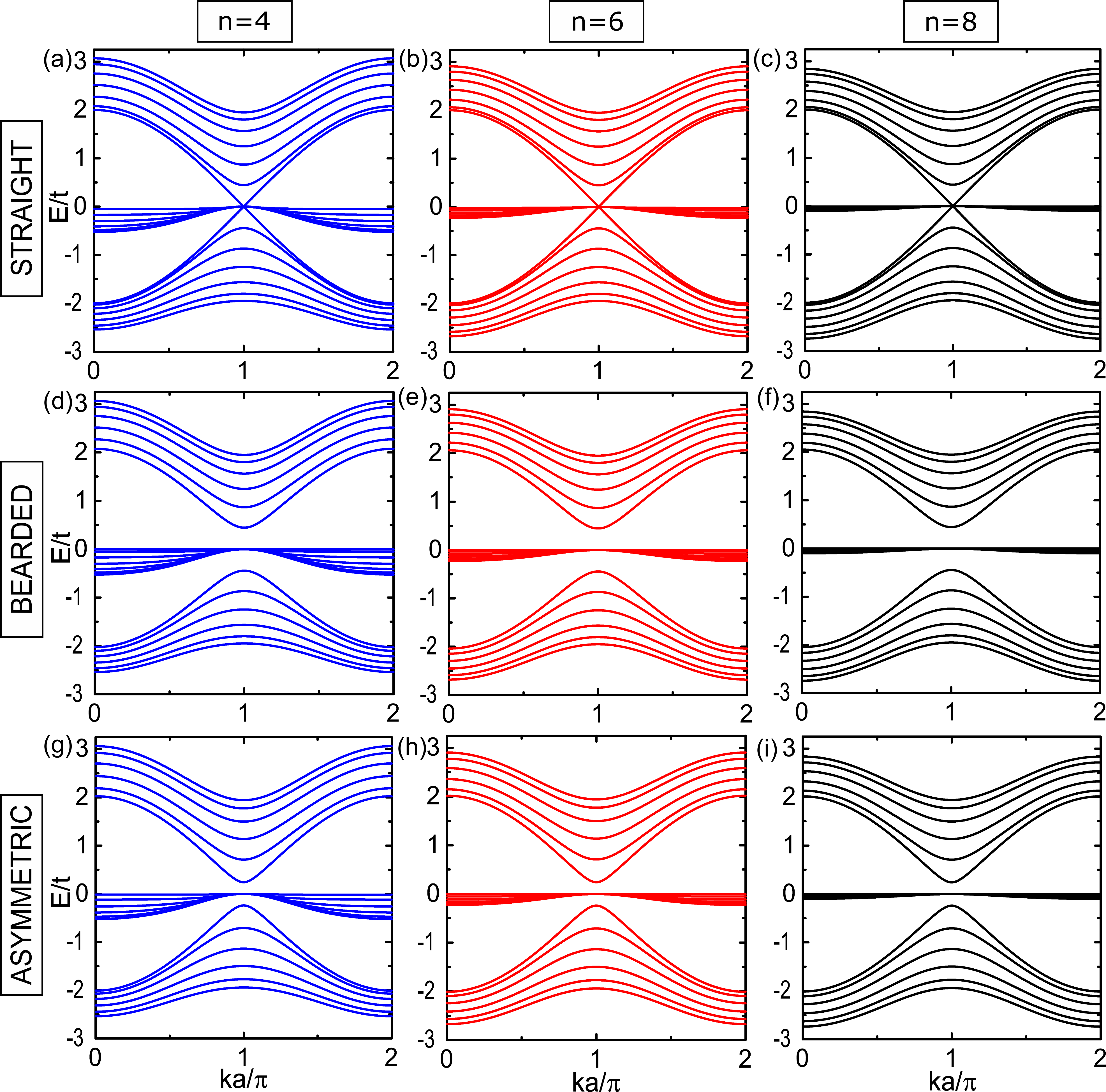}}
\caption{(Color online) Energy spectra of Lieb ($\theta=90^\circ$) nanoribbons with (a)-(c) straight, (d)-(f) bearded, and (g)-(i) asymmetric edges, by taking (a, d, g) $n=4$, (b, e, h) $n=6$, and (c, f, i) $n=8$ in Eq.~\eqref{hopping}. Note the flattening effect of the quasi-flat energy levels by increasing the value of $n$.}
\label{Fig10}
\end{figure}

\begin{figure}[t!]
\centering
{\includegraphics[width=\linewidth]{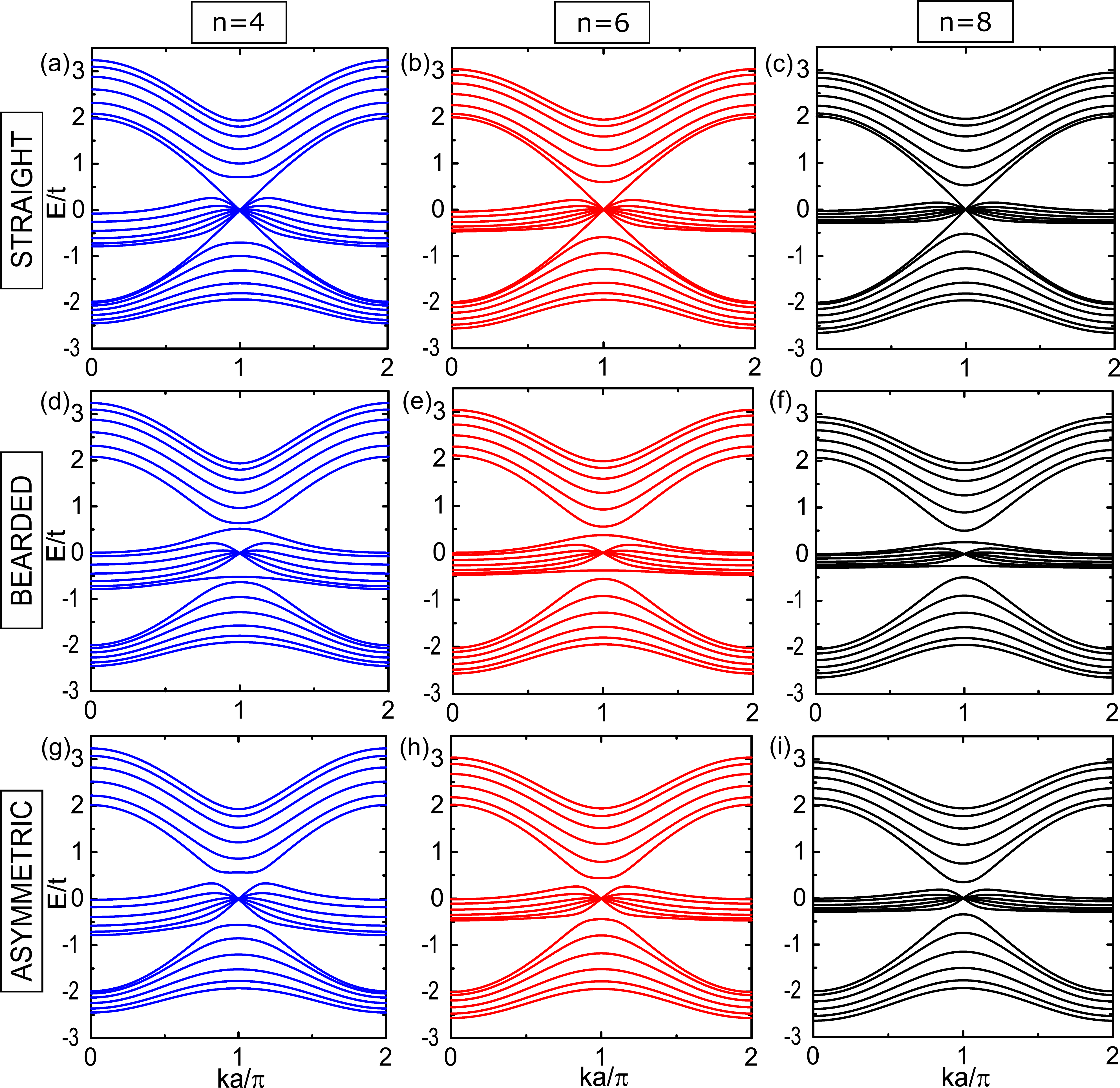}}
\caption{The same as Fig.~\ref{Fig10}, but now for the transition lattice, assuming $\theta=105^{\circ}$.}
\label{Fig11}
\end{figure}

\begin{figure}[t!]
\centering
 {\includegraphics[width=\linewidth]{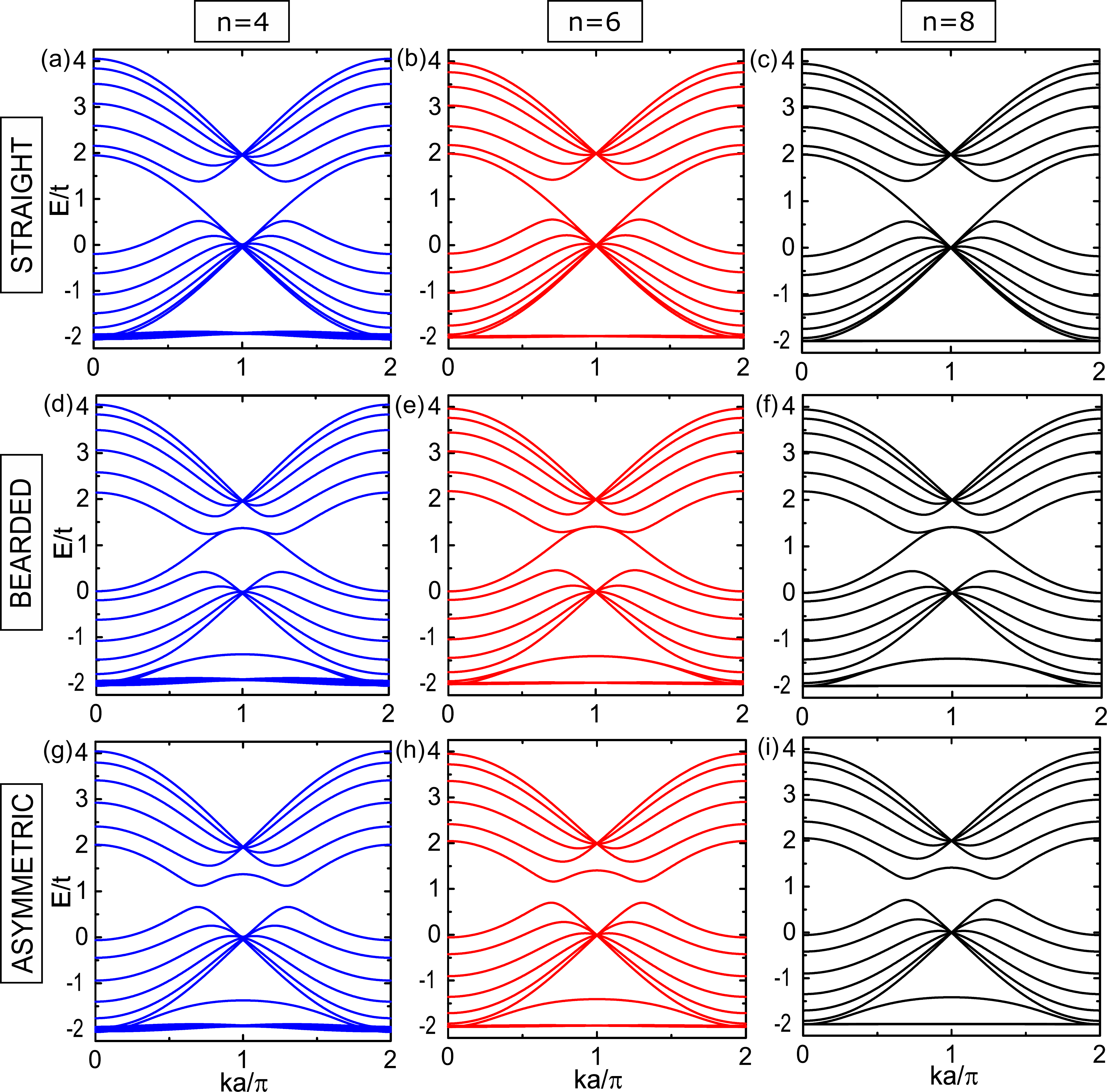}}
\caption{The same as Fig.~\ref{Fig10}, but now for the kagome lattice ($\theta=120^{\circ}$).}
\label{Fig12}
\end{figure}

In order to justify the choice of the parameter $n$ in the hopping energies $t_{i,j}$, as given by Eq.~\eqref{hopping} in Sec.~\ref{sec.II.B}, for the coupling between atoms $i$ and $j$ from sublattices A, B, and C, we present the dispersion relations of Lieb, transition, and kagome nanoribbons in Figs.~\ref{Fig10}, \ref{Fig11}, and \ref{Fig12}, respectively, by taking different values of the $n$-parameter: (left column) $n=4$, (middle column) $n=6$, and (right column) $n=8$; and different edge terminations: (first row) straight edges, (second row) bearded edges, and (third row) asymmetric edges.

Before discussing Figs.~\ref{Fig10}, \ref{Fig11}, and \ref{Fig12}, it is important to analyze the role of the parameter $n$ in the exponential exponent of Eq.~\eqref{hopping}. It works as a strength modulator for the interactions between the connected atomic sites, with its magnitude being modulated by the interatomic distance $a_{ij}$. As expected from Eq.~\eqref{hopping}, larger distances between atoms result in smaller coupling energies. Therefore, the parameter $n$ in Eq.~\eqref{hopping} determines the rate at which the interatomic distance decays. A detailed discussion on the hopping normalization in Lieb, transition, and kagome lattices can be found in Secs.~II, SIV, and SVI of Ref.~[\onlinecite{lima2022}].

From Figs.~\ref{Fig10}, \ref{Fig11}, and \ref{Fig12}, one observes that the value of $n$ controls the quasi-flat subband degeneracy and its non-dispersive character. This is clear for values of $n<8$, where the quasi-flat energy levels exhibit an increase in its bandwidth the smaller the $n$-parameter. Moreover, by increasing $n$, one realizes that the positive (negative) energy levels are energetically squeezed (extend), but no degeneracy break is observed for the levels in the bulk subbands. The breakdown of the quasi-flat subband degeneracy as the value of $n$ decreases can be easily understood by noting that a small $n$ value implies that the hopping energy function [Eq.~\eqref{hopping}] does not decay quickly enough to neglect contributions of sites beyond NN. Therefore, within our generic model described in Sec.~\ref{sec.II}, to capture similar energy levels as the ones reported in the literature \cite{tony2019} with the coexistence of Dirac and quasi-flat energy levels on the Lieb-kagome spectra and obtained with a tight-binding model that approximately takes into account solely NN hoppings, $n=8$ is required. We reiterate that this choice restores the quasi-flat subbands in the energy spectrum of Lieb and kagome nanoribbons, approaching the exactly flat subbands characteristic of the exact NN approximation discussed in Sec.~\ref{Sec.NNN_sites}.

\section*{Acknowledgments}

The authors would like to thank the National Council of Scientific and Technological Development (CNPq) through Universal and PQ programs and the Coordination for the Improvement of Higher Education Personnel (CAPES) of Brazil for their financial support. D.R.C gratefully acknowledges the support from CNPq grants $313211/2021-3$, $437067/2018-1$, $423423/2021-5$, $408144/2022-0$, the Research Foundation—Flanders (FWO), and the Fundação Cearense de Apoio ao Desenvolvimento Científico e Tecnológico (FUNCAP).

\bibliography{References.bib}
\end{document}